\documentclass[twocol]{ametsocV6.2}
\usepackage{mathrsfs,physics}
\usepackage{booktabs,array,tabularx}
\usepackage[hidelinks,hypertexnames=false]{hyperref}

\makeatletter
\def\p@subsection{\thesection}
\newcommand{\appendixlabel}[2]{\begingroup\phantomsection\def\@currentlabel{#1}\label{#2}\endgroup}
\makeatother

\title{Basin-Scale Nonlinear Dynamics of Thermohaline Ocean Circulation.\\ Part I: Theory}

\authors{Conner W. Lester,\aff{a}\correspondingauthor{Conner W. Lester, lesterc@uncw.edu} Dylan E. McNamara,\aff{a} and Till J. W. Wagner\aff{b}}

\affiliation{\aff{a}{Department of Physics and Physical Oceanography, University of North Carolina Wilmington, Wilmington, North Carolina}\\
\aff{b}{Department of Atmospheric and Oceanic Sciences, University of Wisconsin--Madison, Madison, Wisconsin}}

\abstract{Earth's thermohaline ocean circulation
may support multiple equilibria and abrupt transitions. Similar bifurcation structures for different ocean models suggest common underlying dynamics across the model hierarchy. Here we propose a unifying framework from tracer conservation, in which surface fluxes are balanced by dissipative and advective tracer transport. We express this flux balance
through a large-scale tracer gradient and a transport efficiency that measures how strongly circulation displaces the tracer field from its no-flow ``conductive'' state. The flux balance quantifies the transport contributions that sustain each circulation state and give rise to abrupt transitions. In a three-dimensional ocean-basin model, we find nonlinear salinity advection dominates the collapse threshold and linear dissipation dominates the recovery threshold. Temperature has additional dissipation from surface restoring, making thermal fluxes more dissipation-dominated than salinity fluxes. 
The tracer contrasts and boundary layers that a given circulation generates produce distinct transport regimes, approximately quadratic in circulation strength for weak flow and linear for strong flow. With a simple circulation closure, the flux balance reduces to longstanding box-model bifurcation results for the weak and strong flow regimes.
Thus, by expressing thermohaline bifurcations through basin-scale transport, this framework connects box-model insight to the continuous equations. Furthermore, as this framework exposes the transport processes that balance a changing forcing, it provides constraints on a thermohaline ocean state and its trend toward tipping that overturning strength alone does not. In Part~II, we develop flux-balance closures to forecast such tipping points.
}

\begin{document}
\maketitle
\lhead{}

\sigstatement
The Atlantic overturning circulation transports heat and salt between latitudes and may collapse at a critical point as climate changes, yet the dynamics that set this threshold remain uncertain. We show that salt conservation gives a basin-wide dynamical constraint where surface freshwater input is redistributed by advective circulation or dissipated through eddy diffusion. Using a 3D ocean model, we partition advective and diffusive transport near tipping points and decipher how the circulation responds to changing freshwater input. The framework estimates the forcing at which a collapsed state first becomes possible, relates to long-standing box-model results, and shows that overturning strength alone does not reveal how close the circulation is to a tipping point.


\section{Introduction}\label{sec:intro}

Earth's thermohaline ocean circulation is a nonlinear system forced by differential sea-surface heat and freshwater fluxes that establish buoyancy contrasts and shape circulation on global scales \citep{Rahmstorf2002}. This coupled system may support multiple equilibria and abrupt hysteretic transitions between distinct circulation states \citep{Stommel1961,Rahmstorf1995,Weijer2019}. The meridional overturning circulation (MOC) is a key element of the system which transports heat and salt between latitudes and thereby shapes regional and global climate. The Atlantic component of the MOC, the AMOC, transports roughly 1 petawatt of heat northward, so an abrupt change in circulation would reorganize the basin-scale heat budget with consequences for regional to global temperatures, sea ice extent, rainfall, and coastal sea level \citep{Johns2011,Weijer2019}. Furthermore, the AMOC features relatively saline surface waters which support deep-water formation at northern high latitudes \citep{Bryan1986,Weijer2019}. This salinity dependence may make the AMOC susceptible to abrupt transitions as surface heating and freshwater input change under anthropogenic warming \citep{Weijer2019,Dijkstra2026}. 

Numerous modeling efforts, ranging from conceptual to high-order numerical approaches have been directed at understanding the dynamics of thermohaline ocean circulation and its potential for collapse.
Regardless of the level of approximation along this hierarchy, each model framework is physically motivated by the transport dynamics of heat and salt in the ocean. Their basic ingredients are (i) sea-surface fluxes of heat and freshwater, often with (ii) temperature restoring toward atmospheric conditions \citep{Haney1971}, and (iii) advective transport that couples temperature and salinity through a buoyancy-dependent circulation. Some models include additional transport mechanisms through diffusion and gyres.
The key differences along the modeling hierarchy lie in how these processes and their flow-coupling are represented, and in how much spatial and dynamical detail is retained. 

Low-order box models provide
the clearest reduced language for tracing the feedbacks that lead to bifurcations \citep[e.g.,][]{Stommel1961,Cessi1994,Johnson2007,Cimatoribus2014,Wood2019}. In the two-box model of \citet{Stommel1961}, different thermal and haline responses to circulation produce multiple equilibria through the salt-advection feedback: In a thermally driven circulation state, a perturbation that weakens the circulation reduces salt transport to the colder, fresher region, allowing further freshening and additional weakening of the circulation. This feedback can produce an abrupt transition from a temperature-dominated circulation to a salinity-dominated circulation. Subsequent box models have extended this framework to include gyre transport \citep[e.g.,][]{Wood2019,Alkhayuon2019}, wind forcing \citep[e.g.,][]{Johnson2007,Cimatoribus2014}, atmospheric coupling \citep[e.g.,][]{Scott1999}, and additional boxes that tessellate the ocean with increasing detail \citep[e.g.,][]{Cimatoribus2014,Wood2019}. These extensions can change the circulation qualitatively, producing oscillations \citep{RivinTziperman1997,Johnson2007} or even removing the bistable regime altogether \citep{Longworth2005,GuanHuang2008}. 
However, their applicability depends on how well the assumed box structure and buoyancy-flow coupling capture changes in circulation and tracer fields across different thermohaline states \citep{Wunsch2005,deBoer2010}.

Comprehensive general circulation models (GCMs) are found at the other end of the modeling hierarchy, explicitly resolving a wide range of oceanic and atmospheric processes while parameterizing others \citep{Griffies2000,DijkstraGhil2005,Hewitt2020}. 
Low-resolution GCMs have been used to identify multiple AMOC equilibria and explore their bifurcation structure \citep{ManabeStouffer1988,DijkstraWeijer2005,Hawkins2011}. More comprehensive coupled GCMs investigate thermohaline collapse and recovery through transient freshwater hosing experiments \citep{Jackson2023,vanWestenDijkstra2023,vanWesten2024}. The bifurcation structures found in some GCMs are qualitatively similar to those of reduced models lower in the hierarchy \citep{vanWesten2025SaddleNode}. But crucially, GCMs can respond very differently to comparable forcing, both in experiments with increasing atmospheric CO$_2$ \citep{Gregory2005,Baker2025} and in freshwater hosing experiments \citep{Jackson2023,vanWesten2025Eddying}. 
The difficulty lies is determining how changes in heat and salt transport set critical thresholds, and whether the mechanisms isolated in reduced models explain the different responses found across GCMs \citep{Wunsch2005}.

Situated between box models and GCMs in the model hierarchy are intermediate frameworks that solve simplified forms of the spatial thermohaline and momentum equations and have helped connect dynamics and bifurcation structures across the hierarchy \citep{DijkstraWeijer2003,DijkstraGhil2005}. Studies of two-dimensional thermohaline convection, for example, have found asymmetric pole-to-pole circulation states, multiple equilibria with tipping points and cusp bifurcations \citep{ThualMcWilliams1992,CessiYoung1992}, as well as noise-induced transitions \citep{Soons2025}. These non-rotating models, however, omit the geostrophic balances that organize much of the large-scale ocean circulation. Their overturning is therefore governed by different momentum balances, and they are typically studied in parameter spaces far from oceanic conditions \citep{DijkstraGhil2005,Soons2025}. Three-dimensional rotating models retain these geostrophic balances and also resolve the western boundary currents and wind-driven gyres that two-dimensional models must parameterize or omit \citep{DijkstraGhil2005,Wunsch2005}.  Multiple equilibria persist at this level \citep{Bryan1986}, with bifurcations qualitatively similar to those of models across the hierarchy \citep{WeijerDijkstra2001,DijkstraWeijer2003}. 
These commonalities suggest that the conservation laws underlying the system may provide a common framework for understanding the shared nonlinear dynamics and resulting bifurcation behavior. 

Here, we propose such a framework, derived from the temperature and salinity conservation equations. Our aim is to organize the nonlinear dynamical responses found across the model hierarchy through a common description of how circulation redistributes heat and salt in the ocean. The resulting basin-scale thermohaline constraints provide a way to examine both the assumptions of reduced models and the behavior of spatially resolved simulations. We present this work in two parts: Part~I develops the theoretical framework and its immediate implications; Part~II \citep{LesterPart2} proposes and tests methods for using this framework to forecast tipping points.

\begin{figure*}[t!]
\centering
\includegraphics[width=\textwidth]{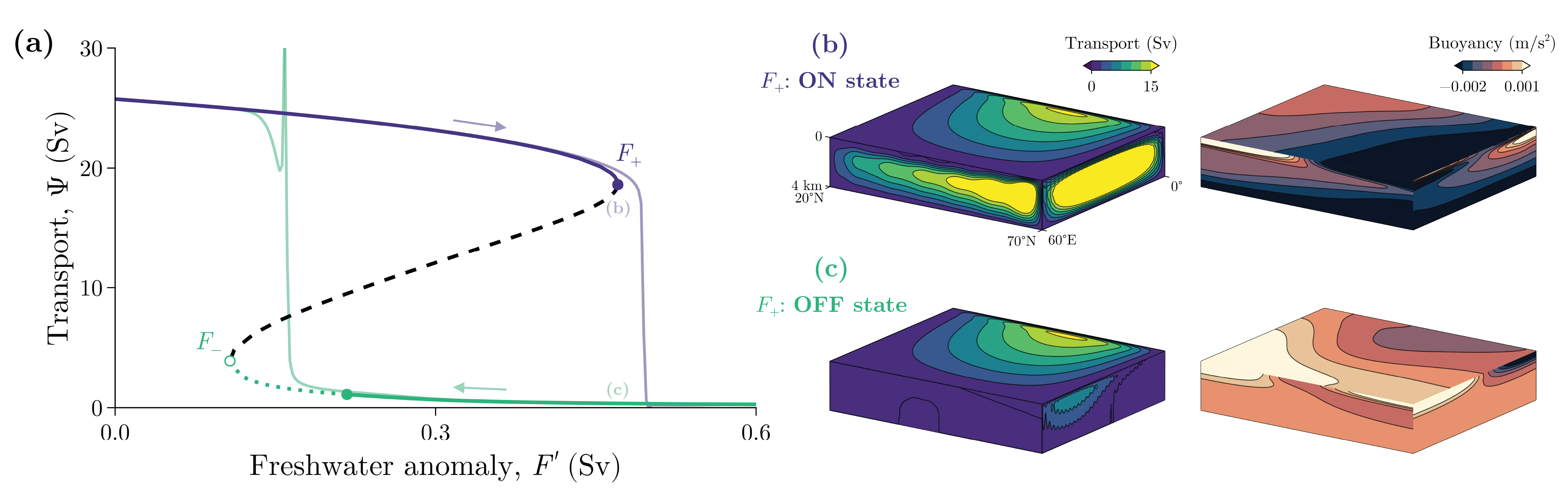}
\caption{Thermohaline bifurcation structure and basin-scale collapse in a 3D rotating basin. (a) Equilibrium overturning transport $\Psi$, the zonally integrated meridional streamfunction at 1000-m depth and $57.5^\circ$N, versus the freshwater forcing anomaly $F'$, relative to a background $\bar F\simeq0.6$~Sv with the same spatial pattern. Solid curves denote stable equilibria and dashed and dotted curves unstable equilibria. $F_+$ and $F_-$ mark the collapse and recovery folds (saddle-node bifurcations), and the solid green circle marks a Hopf bifurcation, beyond which the OFF branch is stable. Faint curves show a transient forcing--recovery cycle with a forcing rate of $0.1$~Sv per 1000 years. (b, c) Streamfunction and buoyancy fields of the coexisting ON and OFF states at the collapse fold $F_+$ (labels in panel a). The horizontal, latitude--depth, and longitude--depth faces show streamfunctions of the depth-, zonally, and meridionally integrated flow, and the corresponding averages of buoyancy, so the fields need not match along shared edges.}

\label{fig1}
\end{figure*}

We begin by describing the general thermohaline dynamics considered here (section~\ref{sec:dynamics}). We then present an idealized 3D numerical model that acts as a testbed for the proposed theoretical framework (section~\ref{sec: numerics}). This framework consists of a basin-scale equilibrium flux balance between imposed surface forcing and the advective and dissipative fluxes that redistribute heat and salt. The flux balance is expressed through domain-integrated variables representing large-scale tracer gradients and advective transport (section~\ref{sec:basin}). 
We examine how circulation shapes boundary layers in the temperature and salinity fields, which in turn affect basin-scale advective tracer transport (sections~\ref{sec:tracer_fields} and~\ref{sec:stirring}). The resulting transport relations are shown to connect to those used in box models and derived in earlier continuous thermohaline models (section~\ref{sec:box_models}). These connections help explain common box-model transport laws and identify limitations of their assumptions. We conclude Part~I by examining how the flux balance distinguishes ocean states with similar overturning strengths but different basin-scale tracer transport and distances from tipping (section~\ref{sec:discussion}).


\section{Thermohaline dynamics}\label{sec:dynamics}

The dynamics of thermohaline circulation are governed by the conservation laws for temperature and salinity anomalies \(T(\boldsymbol{x},t)\) and \(S(\boldsymbol{x},t)\), whose local conservation equations can be expressed as
\begin{align}\label{eq: T,S PDEs}
\partial_t T+\bm u \cdot \bm{\nabla}T 
&=\bm \nabla\cdot(\bm K_T\bm{\nabla}T)+ \mathcal F_{T} - \mathcal R_T T,\nonumber\\
\partial_t S+\bm u \cdot \bm{\nabla}S 
&=\bm \nabla\cdot(\bm K_S\bm{\nabla}S)+ \mathcal F_{S}.
\end{align}
The forcings $\mathcal F_T$ and $\mathcal F_S$, together with the generalized temperature restoring $\mathcal R_T T$, represent surface heat and freshwater fluxes, which broadly warm and salinify the ocean at low latitudes and cool and freshen the ocean toward the poles \citep{Haney1971,WeaverSarachik1991}. For \(X\in\{T,S\}\), the total tracer flux is $\bm J_X= \bm{u}(b)X-\bm K_X\bm{\nabla} X$, where $\bm{u}(b)X$ is the advective flux and $-\bm K_X\bm{\nabla} X$ the anisotropic eddy-diffusive flux, with diagonal eddy diffusivity tensors $\bm K_X$ that may vary in space. Eddy diffusion and temperature restoring are the only forms of tracer dissipation, both having stabilizing effects.  The buoyancy field $b(T,S)$ couples the tracer system through an incompressible flow $\bm u(b)$, driven by sea-surface wind stresses and the evolving buoyancy field. We assume a linear equation of state for buoyancy $b=g(\alpha_TT-\alpha_SS)$. For clarity of presentation in the text, we nondimensionalize buoyancy by gravity $g$ and the tracers $T,S$ by their respective thermal and haline expansion coefficients $\alpha_T,\alpha_S$, such that $b=T-S$. Here we refer to temperature $T$ and salinity $S$ as tracers, scalar quantities that are transported by advection and eddy diffusion.

The conservation laws in Eqs.~\eqref{eq: T,S PDEs} sets the starting point for our flux balance framework derived below. We keep the flow field $\bm u(b)$ as a general buoyancy-coupled field without specifying its momentum dynamics. We represent tracer dissipation (eddy diffusion and temperature restoring) as linear and the forcings $\mathcal F_X$ as independent of the ocean state. Many models include explicit nonlinearities or state couplings in eddy diffusion or surface forcing \citep[e.g.,][]{Griffies2000,Vanderborght2025} that we do not consider here. We discuss the limits of these assumptions in section~\ref{sec:discussion}.

Equations~\eqref{eq: T,S PDEs} form a nonlinear coupled system: surface forcings $\mathcal F_X$ set up tracer and buoyancy fields that drive a circulation $\bm u(b)$; through nonlinear advective transport, this circulation redistributes both tracer fields; the redistributed buoyancy field further alters circulation; and so on until imposed surface fluxes are balanced by advective and diffusive fluxes in steady state. Salinity gradients are set up by a fixed surface freshwater flux field $\mathcal F_S$ and are sensitive to circulation changes. Temperature gradients are set up by $\mathcal F_T-\mathcal R_T T$ and are less sensitive to circulation changes due to the stabilizing restoring term. This asymmetry in the tracer response is the basis of the salt-advection feedback that can produce multiple equilibria and abrupt transitions \citep{Stommel1961, Bryan1986}. 
A strong thermally driven circulation tends to lower salinity gradients that oppose the thermal driving, while temperature restoring helps maintain temperature gradients. 
If the salt-advection feedback becomes sufficiently strong relative to the stabilizing responses of restoring and diffusion, the temperature-driven circulation can tip to an alternative salinity-driven circulation state \citep{Weijer2019}. Such transitions tend to occur as large, ocean basin-scale changes in the circulation and tracer fields \citep[e.g.,][]{vanWesten2024}.

The MOC is a conventional kinematic diagnostic of thermohaline circulation. It measures meridional volume transport, and is used to infer changes in the thermohaline state. Figure~\ref{fig1} shows an equilibrium MOC--freshwater-forcing diagram obtained from a three-dimensional rotating Boussinesq model in which wind- and buoyancy-driven flow $\bm u(b)$ is coupled to the temperature and salinity equations above. The model setup, forcing, and numerical methods are described in section~\ref{sec: numerics} and appendix~\ref{app:numerical_model}. Panel (a) shows an S-shaped equilibrium branch with coexisting ON and OFF states and two fold bifurcations. The transport and buoyancy fields in panels (b, c) show that these ON and OFF states differ substantially through basin-scale reorganization in the overturning and buoyancy fields, while the wind-driven barotropic circulation is the same in both states. The MOC records the change in overturning, but it does not identify how the imposed surface fluxes are balanced by circulation and dissipation mechanisms, or why an equilibrium loses stability or ceases to exist. This missing connection between local temperature and salinity conservation [Eqs.~\eqref{eq: T,S PDEs}] and the basin-scale bifurcation structure motivates the framework of this study.

Our framework derives directly from the temperature and salinity conservation equations [Eqs.~\eqref{eq: T,S PDEs}] on domains that close the heat and salt budgets. For an open domain, the associated heat and salt transports through the open boundaries must instead be retained explicitly in the integrated balances we present below \citep[e.g.,][]{Dijkstra2007}. Here, we consider domains with closed or periodic lateral boundaries and no bottom tracer flux, while retaining sea-surface heat and freshwater fluxes and wind stress as imposed forcing. This closed-system formulation applies to idealized single-hemisphere basin models \citep[e.g., Fig.~\ref{fig1}; ][]{WeaverSarachik1991}, dual-hemisphere Atlantic-like basins connected to a zonally periodic Southern Ocean channel \citep[e.g.,][]{WolfeCessi2014}, and, in principle, GCMs where the global ocean represents the closed domain \citep[e.g.,][]{DijkstraWeijer2005}. 

The flux balance framework derived in section~\ref{sec:basin} from Eqs.~\eqref{eq: T,S PDEs} adds information to the bifurcation structure in Fig.~\ref{fig1} that the MOC cannot provide. Namely, it relates each steady state, stable or unstable, to the shares of advective and dissipative transport that balance the imposed forcing. This constraint reveals how the transport adjusts as the forcing changes, and how transport contributions differ between coexisting ON and OFF states. A low-order closure derived in Part~II allows the flux balance to forecast how these transport contributions change along each branch and thus where tipping will occur.

\section{Numerical simulations}\label{sec: numerics}

We use a three-dimensional Boussinesq model in which temperature and salinity evolve according to Eqs.~\eqref{eq: T,S PDEs} in a rotating basin that represents an idealized North Atlantic. The rectangular Cartesian basin has a flat bottom at a depth of 4~km and extends meridionally from $20^\circ$ to $70^\circ$N, with a zonal width equivalent to $60^\circ$ longitude at the central latitude. The ocean velocity is diagnosed from the instantaneous buoyancy field and imposed wind forcing using a basin-interior geostrophic balance with eddy-diffusion and Rayleigh-drag corrections \citep[e.g.,][]{Huck1999}. The same constant horizontal and vertical coefficients, $\nu_h$ and $\nu_v$, are used for momentum viscosity and for temperature and salinity diffusivity.

The prescribed atmospheric temperature, freshwater flux, and zonal wind stress are zonally uniform with meridional cosine profiles and zero mean. The temperature field is restored to the atmospheric temperature profile with contrast $\Delta T$ at a restoring time per unit depth $\gamma_T^{-1}$. The background and varied anomaly freshwater forcing have amplitudes $\bar F$ and $F'$, respectively, and share the same surface profile. We present the total freshwater forcing amplitude $F=\bar F+F'$ as the freshwater input to the northern half of the basin, reported in sverdrup (1~Sv~$\equiv$~$10^6$~m$^3$~s$^{-1}$) when units are shown. Along each equilibrium branch, the freshwater anomaly amplitude varies while the other parameters remain fixed at their default values (Table~\ref{tab:default_parameters}). These default values are comparable to those of idealized basin models \citep[e.g.,][]{Stommel1948,Huck1999,Arzel2006,Sevellec2009}, with mixing coefficients in the range of tracer diffusivities and momentum viscosities typical of coarse-resolution ocean models \citep[e.g.,][]{Huck1999,Arzel2006,Dijkstra2007,WolfeCessi2014}. For this flat-bottom basin, the depth-averaged circulation is wind driven with wind stress amplitude $\tau_0$ and independent of the buoyancy field, following a Stommel--Munk vorticity balance \citep{Stommel1948,Munk1950}. The barotropic transport streamfunction, shown on the horizontal surface planes of Fig.~\ref{fig1}b,c, therefore remains fixed along each branch.

The system is solved using a Galerkin expansion of the fields into Fourier modes \citep[see, e.g.,][]{deville2002high}. This modal transform retains the conservation constraints of the governing equations and solves them efficiently with a relatively small number of evolving degrees of freedom. As a result, steady-state parameter continuation methods \citep[such as used, for example, by][]{DijkstraWeijer2005} can readily solve for the system's stable and unstable equilibrium branches. Hopf bifurcations occur on the OFF branch near the recovery fold (Fig.~\ref{fig1}), but we focus on bifurcation geometry and mark only the fold bifurcations in subsequent figures. Details of the model formulation and numerical methods are provided in appendix~\ref{app:numerical_model}.

This numerical model does not represent a physically realistic ocean (compared to GCMs, for example), but it retains a wind-driven gyre and a buoyancy-driven overturning circulation that can collapse under freshwater forcing (Fig.~\ref{fig1}). The framework that follows is derived directly from the tracer conservation equations and the stated boundary conditions, independently of the numerical model, and therefore translates to other thermohaline models satisfying the same assumptions. We use these numerical simulations to illustrate thermohaline tipping dynamics and basin-scale flux balances, and to test the transport approximations developed here. In Part~II, we simulate bifurcation branches across a wide range of thermal forcing and mixing parameters to evaluate tipping-point forecasts.

\begin{table}[tb]
\centering
\caption{Key simulation parameters associated with surface forcing and mixing. Values reported are for the default simulation (Fig.~\ref{fig1}).}
\label{tab:default_parameters}
\small
\setlength{\tabcolsep}{3pt}
\renewcommand{\arraystretch}{1.15}
\begin{tabularx}{\columnwidth}{@{}ll>{\raggedright\arraybackslash}X@{}}
\toprule
$\Delta T$ & $20\,{}^\circ\mathrm{C}$ & Imposed temperature contrast \\
$\gamma_T^{-1}$ & $2\,\mathrm{day\,m^{-1}}$ & Thermal restoring time per unit depth \\
$\nu_h$ & $5\times10^3\,\mathrm{m^2\,s^{-1}}$ & Horizontal viscosity/diffusivity \\
$\nu_v$ & $5\times10^{-4}\,\mathrm{m^2\,s^{-1}}$ & Vertical viscosity/diffusivity \\
$\tau_0$ & $0.1\,\mathrm{N\,m^{-2}}$ & Wind stress amplitude \\
\bottomrule
\end{tabularx}
\end{table}

\section{Basin-scale dynamics}\label{sec:basin}

To derive basin-scale balances from Eqs.~\eqref{eq: T,S PDEs}, we first write the temperature and salinity equations in a common tracer form. We define the nonlinear advection operator as \(\mathcal A(b)X=\bm u(b)\boldsymbol{\cdot}\bm\nabla X\) and the total linear dissipation operator as \(\mathcal D_X X=-\bm\nabla\boldsymbol{\cdot}(\bm K_X\bm\nabla X)+\mathcal R_X X\). The latter combines eddy diffusion and thermal restoring, with \(\mathcal R_S=0\). The derivation below applies to either tracer \(X\). For clarity of presentation, we suppress tracer labels, for example \(\mathcal D_X\to\mathcal D\), \(\mathcal F_X\to\mathcal F\), etc. We restore \(T\) and \(S\) labels where the two tracers must be distinguished. Unless otherwise stated, unlabeled quantities in the figures refer to salinity, with $F=F_S$ the freshwater forcing amplitude. With this convention, the $X\in \{T,S\}$ conservation equations [Eqs.~\eqref{eq: T,S PDEs}] can be written as
\begin{equation}\label{eq: operator PDEs}
\partial_tX+\mathcal A(b)X+\mathcal D X=\mathcal F.
\end{equation}

In steady state, advection and dissipation jointly balance the imposed forcing, $\mathcal A(b)X+\mathcal D X=\mathcal F$. This entails that the tracer field $X(\bm x)$ needed to maintain this balance depends on the circulation state $\bm u(b)$. A useful measure of the effects of circulation on the tracer field is the ``conductive'' reference state $X_c$ that balances surface forcing through dissipation alone:
\begin{equation}\label{eq: X_c}
  \mathcal D X_c = \mathcal F.
\end{equation}
Comparing the realized field \(X\) with the conductive field \(X_c\) then reveals how circulation changes the tracer distribution required to balance the forcing in steady state. Comparisons with a conductive reference are commonly used in similar fluid systems to measure the effects of stirring, meaning the advective redistribution of a tracer by the flow, on continuously forced tracers \citep{Thiffeault2012,ShawThiffeaultDoering2007} and the contribution of circulation to heat transport in horizontal convection \citep{Rocha2020}. Using the conductive reference state, the full tracer equation can be written as
\begin{equation}\label{eq: PDE w X_c-X}
\partial_tX+\mathcal A(b)X =\mathcal D(X_c-X).
\end{equation}
The difference \(X_c-X\) measures the stirring-induced departure of the tracer field $X$ from the conductive reference $X_c$. At equilibrium, the dissipation of this difference balances the advective redistribution of the realized tracer field.

\subsection{Flux balance}\label{sec:flux_balance}

We define basin-scale variables by the normalized volume average $\langle f\rangle=\int f\,\dd V$ and inner product $\langle f,g\rangle=\int fg\,\dd V$, with $\int\dd V=1$. Averaging the tracer equations over the domain gives the temperature and salinity budgets $\partial_t\langle T\rangle = \langle \mathcal F_T\rangle - \langle \mathcal R_T T\rangle,$ and $\partial_t\langle S\rangle = 0.$ These budgets contain no explicit advective or diffusive transport terms, since internal advective and diffusive redistribution do not change total tracer content. To retain information about the tracer distribution, we multiply the tracer equation by $X$ and take the volume average, obtaining the variance balance
\begin{equation}\label{eq: variance}
  \tfrac12\partial_t\langle X^2\rangle = \langle X,\mathcal D (X_c-X)\rangle.
\end{equation}
Advection has no direct effect on variance production due to incompressibility in the closed system considered here, which ensures \(\langle X,\mathcal A(b)X\rangle=0\). Advection nevertheless changes the spatial structure of \(X\), with indirect advective effects measured by \(X_c-X\). To obtain an explicit measure of advective redistribution, we project the tracer equation onto the conductive field. This gives a basin-scale balance with separate forced, advective, and dissipative contributions.

We first separate the conductive field into its spatial pattern and amplitude:
\begin{equation}\label{eq: X_c = F_X e^X_c}
  X_c(\bm x) = F e_c(\bm x),\qquad F=\langle e_c,\mathcal F\rangle,
\end{equation}
where \(F\) is the imposed surface-forcing amplitude and \(e_c(\bm x)\) is the dissipation-normalized conductive field, \(\langle e_c,\mathcal D e_c\rangle=1\). Along each simulated bifurcation branch, the spatial forcing pattern and linear dissipation operator remain fixed, so the conductive pattern \(e_c=X_c/F\) remains fixed as the freshwater-forcing amplitude \(F=F_S\) varies.

Recognizing that $F$ represents the basin-scale surface flux magnitude, we can project the steady tracer equation [Eq.~\eqref{eq: operator PDEs}] onto the conductive basis \(e_c(\bm x)\). This gives the basin-scale flux balance
\begin{equation}\label{eq: F = J^diss + J^adv}
  F=J_{\rm diss}+J_{\rm adv},
\end{equation}
where the advective and dissipative contributions are
\begin{align}\label{eq: Jadv, Jdiss}
J_{\rm adv}
&=\langle e_c,\mathcal A(b)X\rangle =-
\int
\bm{u}X\cdot\bm{\nabla} e_c\,\dd V,\\
J_{\rm diss}
&=\langle e_c,\mathcal D X\rangle
=
\int
\big[\bm K \bm{\nabla} X\cdot \bm{\nabla} e_c + e_c \mathcal R X\big]\,\dd V. \nonumber
\end{align}
The advective and diffusive integrals measure tracer transport across contours of the conductive field, weighted by \(\bm\nabla e_c\). For temperature, \(J^T_{\rm diss}\) includes additional dissipation from surface restoring.
 
The projected flux balance of Eq.~\eqref{eq: F = J^diss + J^adv} turns the high-dimensional tracer equations [Eqs.~\eqref{eq: T,S PDEs}] into a basin-scale transport budget and provides a different interpretation of the MOC--freshwater-forcing diagram in Fig.~\ref{fig1}: At equilibrium, the imposed salinity forcing is balanced by the sum of dissipative $J_{\rm diss}$ and advective $J_{\rm adv}$ tracer transport, whose relative contributions vary across the bifurcation branch. The MOC provides one possible diagnostic of the circulation but cannot distinguish the tracer transport contributions that balance a given forcing $F$. 

The steady state variance balance [Eq.~\eqref{eq: variance}] reveals the orthogonality condition 
\begin{equation}\label{eq: orthogonality}
\langle X,\mathcal D (X_c-X)\rangle=\langle X, X_c-X\rangle_{\mathcal D}=0,
\end{equation}
which states that the realized field \(X\) is orthogonal to the circulation-induced displacement \(X_c-X\) in dissipation-weighted space. This orthogonality condition relates realized tracer states to displaced advective states and selects the state-space geometry of the system through the dissipation-weighted inner product $\langle f,g\rangle_{\mathcal D}=\langle f,\mathcal D g\rangle$, in which the tracer dissipation rate is the squared norm $\|X\|^2_{\mathcal D}=\langle X,\mathcal D X\rangle$. Writing $X_c=X+(X_c-X)$, orthogonality gives the Pythagorean identity
\begin{equation}\label{eq: Pythagorean}
\underbrace{\left\|X_c\right\|_{\mathcal D}^{2}}_{F^2}
=
\underbrace{\left\|X\right\|_{\mathcal D}^{2}}_{FJ_{\mathrm{diss}}}
+
\underbrace{\left\|X_c-X\right\|_{\mathcal D}^{2}}_{FJ_{\mathrm{adv}}},
\end{equation}
Here $\|X_c\|^2_{\mathcal D}=F^2$ by the definition $X_c=Fe_c$ and normalization of $e_c$ [Eq.~\eqref{eq: X_c = F_X e^X_c}]. In steady state $\mathcal A(b)X=\mathcal D(X_c-X)$ from Eq.\eqref{eq: PDE w X_c-X}, thus the fluxes [Eq.~\eqref{eq: Jadv, Jdiss}] are $FJ_{\rm diss}=\langle X_c,X\rangle_{\mathcal D}$ and $FJ_{\rm adv}=\langle X_c,X_c-X\rangle_{\mathcal D}$, which orthogonality reduces to $\|X\|^2_{\mathcal D}$ and $\|X_c-X\|^2_{\mathcal D}$. Thus, dividing Eq.\eqref{eq: Pythagorean} by $F$ recovers the basin-scale flux balance [Eq.~\eqref{eq: F = J^diss + J^adv}]. The dissipative flux contribution is therefore proportional to the squared norm of the realized field, while the advective flux contribution is proportional to the squared departure from the conductive reference.

The variance relation above [Eq.~\eqref{eq: Pythagorean}] allows us to constrain \(J_{\rm adv}/J_{\rm diss}\) explicitly as the square of the dimensionless transport efficiency \(q\), representing the advection-induced displacement from the conductive state to the realized state. Dividing Eq.~\eqref{eq: Pythagorean} by $\|X\|^2_{\mathcal D}$, allows us to define the transport efficiency by
\begin{equation}\label{eq: q}
q^2\equiv\frac{\|X_c-X\|^2_{\mathcal D}}{\|X\|^2_{\mathcal D}}
=\frac{J_{\rm adv}}{J_{\rm diss}}.
\end{equation}
The transport efficiency can also be written as \(q=\|\mathcal A(b)X\|_{\mathcal D^{-1}}/\|\mathcal D X\|_{\mathcal D^{-1}}\), representing a P\'eclet-number scale that measures tracer advection relative to dissipation, with \(\| f\|^2_{\mathcal D^{-1}}=\langle f, \mathcal D^{-1} f\rangle\). The squared transport efficiency $q^2$ is a basin-scale analog of a dissipation-normalized box-model advective-transport rate, as discussed in section~\ref{sec:box_models}. When \(q^2<1\) the flux balance is dominated by dissipative transport, while \(q^2>1\) corresponds to advection-dominated transport. Here, we can also identify the fixed-flux Nusselt number \(\mathrm{Nu}=\|X_c\|^2_{\mathcal D}/\|X\|^2_{\mathcal D}=1+q^2\), which measures the reduction in tracer dissipation relative to the conductive reference as tracers are redistributed by circulation \citep{Rocha2020}. For \(\mathrm{Nu}\sim1\), the tracer field remains close to its conductive state and dissipation balances most of the forcing. For \(\mathrm{Nu}\gg1\), circulation balances most of the forcing, allowing the imposed flux to be maintained with much smaller tracer gradients compared to the conductive state. 

Isolating the transport efficiency $q$ in the flux balance [Eqs.~\eqref{eq: F = J^diss + J^adv}] gives $F=J_{\rm diss}(1+q^2)$. Within dissipation-weighted space, the dissipative flux $J_{\rm diss}=\langle e_c,X\rangle_{\mathcal D}$ represents the magnitude of the realized state $X$ in the primary forcing direction $e_c$. This motivates the reinterpretation of the dissipative flux in dissipation-space as
\begin{equation}\label{eq: G}
  G\equiv\langle e_c,X\rangle_{\mathcal D}=J_{\rm diss},
\end{equation}
where \(G\) represents the effective basin-scale tracer gradient aligned with the conductive pattern. Here, \(e_c\) describes the normalized tracer pattern maintained by the imposed forcing in the absence of advective redistribution, such as the salinity gradients associated with meridionally varying freshwater fluxes. Thus \(G\) measures how much of this forcing pattern remains in the realized field after redistribution by circulation. In the limit of no advective redistribution, \(X\to X_c\) and \(G\to\langle e_c,X_c\rangle_{\mathcal D}=F\), recovering the conductive value. The steady variance balance also identifies $G/F=\|X\|^2_{\mathcal D}/\|X_c\|^2_{\mathcal D}=\text{Nu}^{-1}\leq1$ reinforcing that advective redistribution of the tracer field reduces its gradients relative to the imposed forcing.

\begin{figure*}[t!]
\centering
\includegraphics[width=\textwidth]{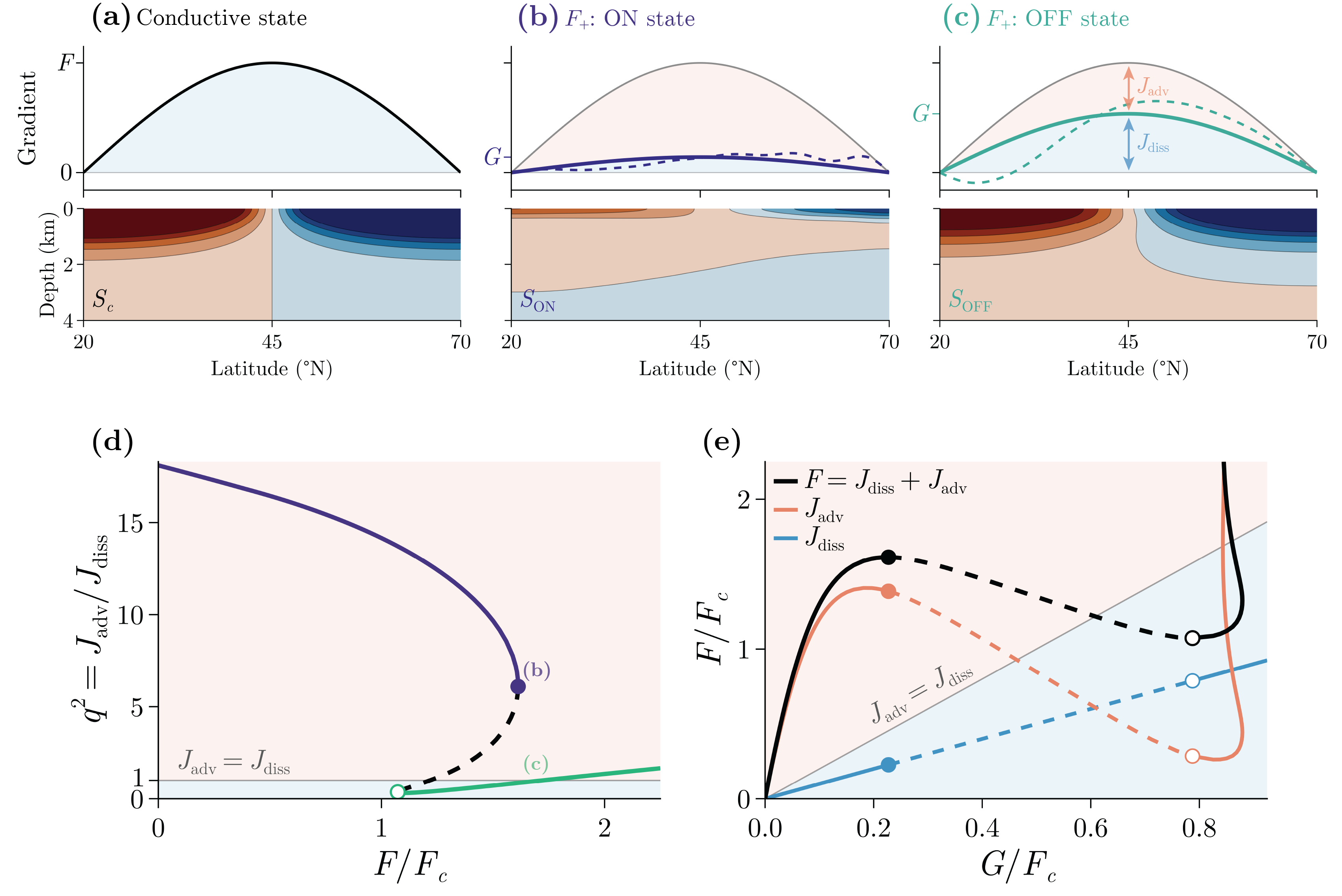}
  \caption{Basin-scale salinity flux balance derived from the same bifurcation simulation shown in Fig.~\ref{fig1}. Panels (a--c) show zonally averaged salinity fields and meridional gradient schematics for the conductive field $S_c$ (a) and realized field $S$ at the ON state (b) and OFF state (c) of collapse freshwater forcing threshold $F_+$. The zonal-mean fields show \(S_c/F=e_c\) and \(S/F\) on a common color scale. Dashed curves show weighted-average meridional gradients of the realized salinity fields below. Projecting these profiles onto the conductive-gradient pattern (a) gives the solid colored curves in (b,c), whose amplitudes \(G\) are reduced relative to the conductive reference amplitude \(F\) due to advective redistribution. Blue and red shading illustrate the basin-scale dissipative contribution \(J_{\rm diss}=G\) and advective contribution \(J_{\rm adv}=F-G\), respectively. Panel (d) shows the transport efficiency $q^2=J_{\rm adv}/J_{\rm diss}$ versus freshwater forcing $F$ (rescaled by conductive forcing scale $F_c$; Eq.~\eqref{eq: Fc}). Shaded regions connect to gradient shading in panels (a--c) with red shade for $J_{\rm adv}>J_{\rm diss}$ and blue for $J_{\rm adv}<J_{\rm diss}$. Panel (e) decomposes the flux balance in (d) showing the total flux $F=J_{\rm diss}+J_{\rm adv}$ (black) and its components $J_{\rm diss}$ (blue) and $J_{\rm adv}$ (orange) as a function of the gradient scale $G$ (with all fluxes rescaled by $F_c$). Advection-dominated flux balance occurs where $J_{\rm adv}>J_{\rm diss}$ (orange line $>$ blue line) and dissipation-dominated balance where $J_{\rm adv}<J_{\rm diss}$ (orange line $<$ blue line). The grey line marks $J_{\rm adv}=J_{\rm diss}$, where $F=2G$, and the red and blue shading, as in (d), applies to the total flux $F$ (black). Collapse $F_+$ (solid circles) and recovery $F_-$ (open circles) fold points are marked.}
  \label{fig2}
\end{figure*}

With this, the complete basin-scale flux balance can be expressed as
\begin{equation}\label{eq: F_X = G_X(1+q_X^2)}
F=J_{\rm diss}+J_{\rm adv}=G(1+q^2).
\end{equation}
This represents a basin-scale transport constraint in terms of the large-scale gradient \(G\) and transport efficiency \(q\)---both identified as basin-scale state variables. The ratio \(G/F=1/(1+q^2)\) gives the fraction of the imposed forcing balanced by dissipation, which decreases with increasing advective redistribution. 

This flux balance is a projection onto the total conductive field \(X_c(\bm x)=\bar X_c(\bm x)+X_c'(\bm x)\), whose background and anomaly parts are the conductive responses to the forcings \(\bar{\mathcal F}(\bm x)\) and \(\mathcal F'(\bm x)\). In our simulations these have the same spatial pattern, so $e_c$ is fixed, and the balance in terms of the anomaly amplitude is $F'=G'(1+q^2)$, with $G'=(F'/F)G$. When the anomaly differs in pattern from the background, as in North Atlantic hosing experiments in GCMs, the flux balance still holds, but $e_c$ changes with the anomaly amplitude, so $q$ changes partly because the reference pattern changes, not only because the circulation does. Appendix~\ref{app:background_anomaly} gives one way to retain a fixed reference pattern, using a passive tracer forced only by the anomaly.

Figure~\ref{fig2} illustrates the basin-scale variables and flux balance for the bifurcation simulation presented in Fig.~\ref{fig1}. Panels (a--c) show zonally averaged salinity fields for the conductive state (a) and ON (b) and OFF (c) states at the collapse forcing $F_+$. Panels (d) and (e) in Fig.~\ref{fig2} recast the MOC--freshwater-forcing bifurcation diagram from Fig.~\ref{fig1}a in terms of the haline transport efficiency $q^2=J_{\rm adv}/J_{\rm diss}$ and flux contributions to \(F=J_{\rm diss}+J_{\rm adv}\). This highlights bifurcation regimes dominated by advection $J_{\rm adv}>J_{\rm diss}\ (q^2>1)$ and dissipation $J_{\rm adv}<J_{\rm diss}\ (q^2<1)$. The large-scale salinity variables \(G=G_S\) and \(q=q_S\) presented in Fig.~\ref{fig2} are implicitly coupled to their temperature counterparts \(G_T\) and \(q_T\) through the buoyancy-dependent circulation, as discussed below. The salinity-flux partitions in panels (d) and (e) show an advection-dominated flux balance along the ON branch where advective transport exceeds dissipative transport ($q^2>1$), including at the collapse fold \(F_+\). By contrast, the recovery fold \(F_-\), which marks the lower forcing limit of the bistable regime, and neighboring equilibrium states are dominated by dissipation ($q^2< 1$). Even though wind-driven circulation and its associated advective effects persist at recovery $F_-$, the total advective contribution is smaller than dissipation, as measured by $q^2=J_{\rm adv}/J_{\rm diss}$. We note that for large \(F\), advection also dominates on the OFF branch---a feature not apparent from the overturning diagnostic in Fig.~\ref{fig1}.

\subsection{Onset of bistability and salt-advection feedback}\label{sec:bistability}

The recovery forcing $F_-$ marks the onset of bistability---where ON and OFF states can exist simultaneously. Identifying the bistable regime sheds light on the fundamental dynamics of the system and carries additional importance because stochastic fluctuations can drive transitions between these states before the deterministic collapse threshold is reached \citep{Cessi1994,Soons2025}. To understand what sets the scale of the recovery forcing $F_-$, and thus the onset of bistability, we turn to the opposing effects of temperature and salinity on buoyancy-driven circulation. 

\begin{figure}[t!]
\centering
\includegraphics[width=\columnwidth]{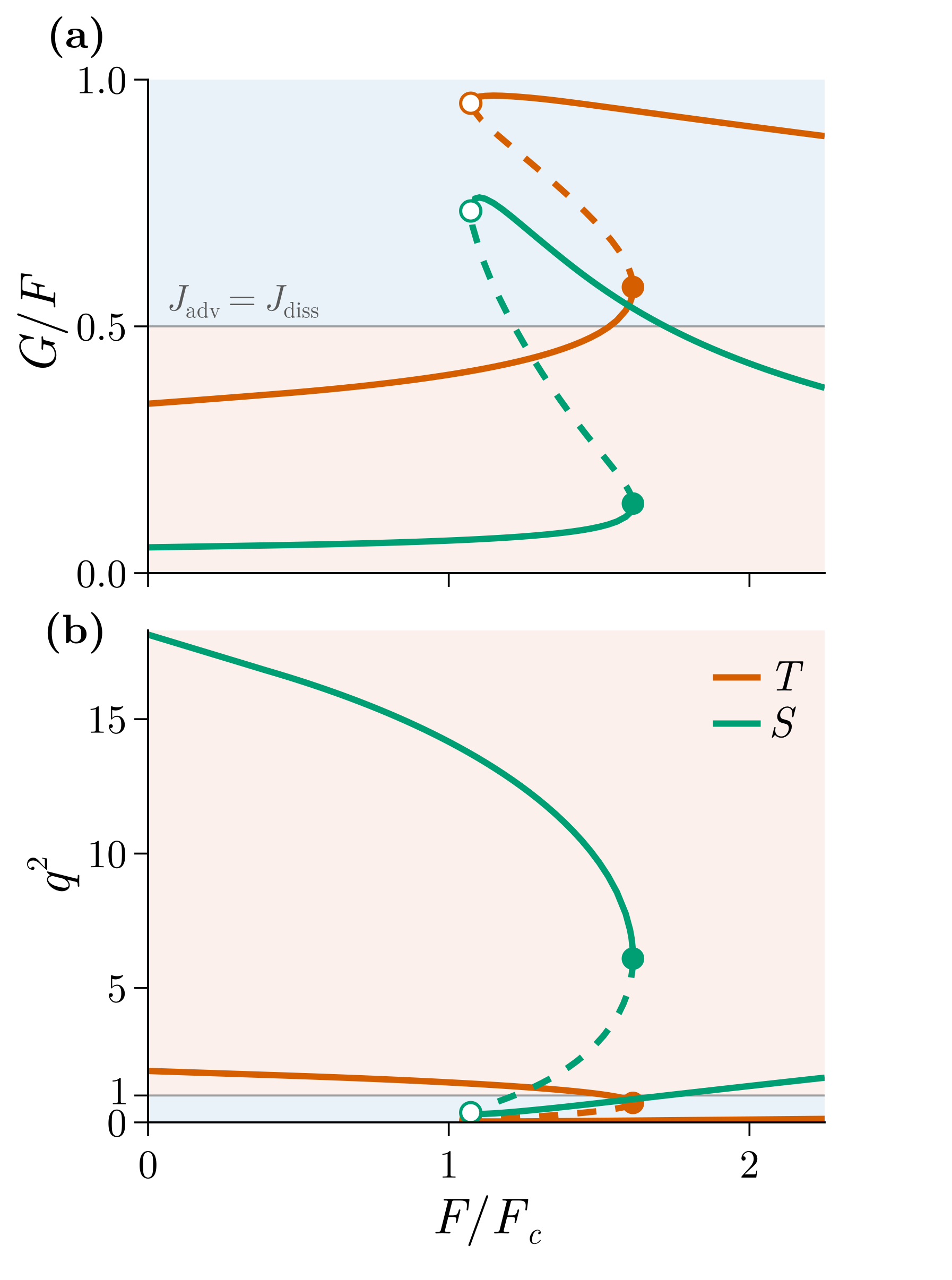}
\caption{Temperature (orange) and salinity (green) flux balances for the same bifurcation simulation shown in Figs.~\ref{fig1} and~\ref{fig2} (Table~\ref{tab:default_parameters}). (a) Fraction of the imposed forcing balanced by dissipation, or equivalently the basin-scale tracer gradient relative to the imposed forcing, $G_X/F_X=1/(1+q_X^2)$, and (b) transport efficiency $q_X^2=J^X_{\rm adv}/J^X_{\rm diss}$ for $X\in\{T,S\}$, versus freshwater forcing $F=F_S$ (rescaled by $F_c$). The thermal forcing amplitude $F_T$ is fixed along the branch, and the thermal dissipative flux $J^T_{\rm diss}$ includes surface restoring. Grey lines mark $J_{\rm adv}=J_{\rm diss}$, with red shading for $J_{\rm adv}>J_{\rm diss}$ and blue for $J_{\rm adv}<J_{\rm diss}$, as in Fig.~\ref{fig2}. Dashed curves show the unstable branch between the folds. Collapse $F_+$ (solid circles) and recovery $F_-$ (open circles) fold points are marked.}
\label{fig3}
\end{figure}

The conductive buoyancy field \(b_c=T_c-S_c\) provides a reference for the buoyancy-driven flow in the dissipation-dominated regime around the recovery fold (Figs.~\ref{fig2}d, e and~\ref{fig3}). For example, an ocean initialized at rest with \(T=T_c\) and \(S=S_c\) and no wind forcing has a motionless equilibrium if \(b_c=0\) everywhere. In general, the conductive fields \(T_c(\bm x)\) and \(S_c(\bm x)\) can differ spatially because of different surface forcing profiles \(\mathcal F_T(\bm x),\ \mathcal F_S(\bm x)\) or different thermal and haline boundary-layer structures arising from different eddy diffusivities. However, even when these fields differ spatially, we can identify the freshwater forcing amplitude at which they most closely compensate. Holding \(T_c\), \(e_c^S\), and \(\mathcal D_S\) fixed, we define the characteristic conductive freshwater forcing \(F_c^S\) as the amplitude that minimizes \(\|b_c(F_S)\|_{\mathcal D_S}^2 =\|T_c-F_S e_c^S\|_{\mathcal D_S}^2\):
\begin{equation}\label{eq: Fc}
  F^S_c = \langle e^S_c,T_c\rangle_{\mathcal D_S}.
\end{equation}
The analogous conductive forcing for temperature is $F^T_c = \langle e^T_c,S_c\rangle_{\mathcal D_T}$. The projected conductive buoyancy is therefore \(\langle e_c^S,b_c\rangle_{\mathcal D_S}=F^S_c-F\), with $F$ as defined in Eq.~\eqref{eq: X_c = F_X e^X_c}. For simplicity, we write \(F_c=F_c^S\). This characteristic forcing marks the conductive temperature and salinity compensation point and provides a natural flux scale for the system. For a forcing anomaly whose spatial pattern differs from the background, appendix~\ref{app:background_anomaly} gives the analogous compensation scale.

In the salinity forcing direction, the conductive buoyancy contrast is temperature dominated for \(F<F_c\) and salinity dominated for \(F>F_c\). When the conductive temperature and salinity fields have similar spatial structures, their near cancellation around \(F_c\) leaves weak conductive buoyancy gradients, providing a reference for a weak buoyancy-driven circulation state. Because $F_c$ marks the point where a weak circulation state may form, this suggests that the recovery forcing should be of order \(F_c\), giving the estimate \(F_-\sim F_c\) for the lower forcing limit of bistability, consistent with the dissipation-dominated recovery fold in Fig.~\ref{fig2}d,e. Spatial differences between \(T_c\) and \(S_c\), together with wind-driven transport, can shift recovery away from \(F_c\). We note that an alternative and possibly more dynamically meaningful compensation scale could be defined by minimizing total advection of the conductive field \(\mathcal A(b_c)e_c^S\) by weak $b_c(F_S)$-driven flow plus any wind-driven advection. Dynamical connections between conductive forcing scales such as \(F_c\) and the bistability onset \(F_-\) also appear in earlier thermohaline models, such as diffusive extensions of the \citet{Stommel1961} box model \citep{Cessi1994,Longworth2005}, two-dimensional Boussinesq simulations \citep{ThualMcWilliams1992}, and asymptotic analysis of continuous thermohaline circulation \citep{CessiYoung1992}.

Note that an established ON circulation can nevertheless persist for $F>F_c$, because advective redistribution reduces the opposing salinity gradients, while surface temperature restoring maintains the thermal gradients that drive the flow (Fig.~\ref{fig3}a). Both tracers share the same flow and eddy diffusivities, but restoring adds to the temperature dissipation, giving a smaller thermal transport efficiency, $q_T<q_S$, along the entire default bifurcation branch (Fig.~\ref{fig3}b). The added dissipation is large enough that the thermal flux balance is dissipation dominated ($q_T^2<1$) at the collapse fold, while the haline flux balance is strongly advection dominated ($q_S^2>1$). However, at fixed $F$, a weaker circulation with a smaller haline transport efficiency $q$ corresponds to a larger equilibrium salinity gradient, $G=F/(1+q^2)$. The larger salinity contrast can, in turn, weaken circulation and salt transport, reinforcing the initial reduction through the salt-advection feedback. The strength of this feedback and the location of the collapse tipping point depend on how circulation redistributes temperature and salinity away from their conductive state, and how that redistribution couples back to the buoyancy driven flow.

\section{Tracer fields and buoyancy coupling}\label{sec:tracer_fields}

We now relate the spatial tracer fields to the basin-scale variables \(G\) and \(q\). This connects the flux balance to the tracer contrasts and boundary layers that contribute to buoyancy and drive circulation.

We decompose each tracer into components parallel and orthogonal to its conductive pattern: \(X=X_\parallel+X_\perp\), with \(\langle e_c,X_\perp\rangle_{\mathcal D}=0\). Since \(\|e_c\|_{\mathcal D}=1\) and \(G=\langle e_c,X\rangle_{\mathcal D}\), the parallel component is \(X_\parallel=Ge_c\). The steady variance balance [Eq.~\eqref{eq: Pythagorean}] gives \(\|X\|_{\mathcal D}^2=FG\), so the orthogonal component satisfies $\|X_\perp\|_{\mathcal D}^2 =\|X\|_{\mathcal D}^2-G^2 =FG-G^2 =G^2q^2,$ using \(F=G(1+q^2)\). Defining the redistribution field \(Q=X_\perp/G\) where $\langle e_c,Q\rangle_{\mathcal D}=0$, the tracer field therefore takes the form
\begin{equation}\label{eq: X field}
  X=G(e_c+Q),\qquad
  \|Q\|_{\mathcal D}=q.
\end{equation}

In this representation, \(G\) is the amplitude of the component aligned with the conductive pattern, while \(Gq\) is the amplitude of the orthogonal redistribution generated by advection. Thus \(q\) measures the relative magnitude of the redistribution field to the conductive field in dissipation-weighted space: the conductive component $e_c$ dominates for \(q<1\), while the advective redistribution component $Q$ dominates for \(q>1\). Circulation changes both components, reducing the conductive amplitude from \(F\) to \(G=F/(1+q^2)\) and generating the redistribution field \(GQ\). This redistribution includes boundary layers and other tracer contrasts generated by advective stirring.

Restoring the tracer labels and writing \(Q_X=q_Xe_q^X\), where
\(\|e_q^X\|_{\mathcal D_X}=1\), the buoyancy field becomes
\begin{align}\label{eq: b field}
  b = \frac{F_T}{1+q_T^2}\Big(e_c^T+q_Te_q^T\Big)
    - \frac{F_S}{1+q_S^2}\Big(e_c^S+q_Se_q^S\Big).
\end{align}
Both the conductive and redistributed components of temperature and salinity contribute to the buoyancy that drives circulation. In turn, the flow \(\bm u(b)\) determines the redistribution amplitudes \(q_T\) and \(q_S\) and their spatial patterns \(e_q^T\) and \(e_q^S\). This coupling allows circulation to generate tracer contrasts and boundary layers that modify the buoyancy field and feed back on the flow. To understand how this coupling enters the basin-scale flux balance, we examine how circulation-induced stirring generates the redistribution fields and sets the transport efficiencies \(q_T\) and \(q_S\).

\section{Stirring response and transport regimes}\label{sec:stirring}

We use stirring to refer to mechanical advective redistribution of a tracer field \citep{Thiffeault2012}. Stirring generates additional tracer contrasts and boundary layers relative to the conductive state (e.g., Fig.~\ref{fig2}a-c). The resulting spatial structure can affect how tracer transport scales with circulation strength \citep{ShawThiffeaultDoering2007}. Below, we examine this relationship through the transport efficiency \(q\).

To express the stirring response of a tracer field relative to the conductive reference, we apply \(\mathcal D^{-1}\) to the steady tracer equation, giving
\begin{equation}\label{eq: X + PX = Xc}
  X+\mathcal P(b)X=X_c,\qquad
  \mathcal P(b)=\mathcal D^{-1}\mathcal A(b).
\end{equation}
The dimensionless operator \(\mathcal P\) compares advection with tracer dissipation and is therefore an operator analogue of a P\'eclet number. This operator conserves tracer variance in dissipation space, \(\langle X,\mathcal PX\rangle_{\mathcal D} =\langle X,\mathcal AX\rangle=0\). From Eq.~\eqref{eq: X + PX = Xc}, the tracer state is determined by the nonlinear inversion \(X=[1+\mathcal P(b)]^{-1}X_c\), which can be expanded as follows:
\begin{align}\label{eq: X series}
X=X_c-\mathcal PX_c+\mathcal P^2X_c-\mathcal P^3X_c +\cdots.
\end{align}
The successive terms represent repeated redistribution of the conductive pattern \(X_c\) by the circulation. The first stirring term $\mathcal PX_c$ moves the field away from the conductive direction \(\langle e_c,\mathcal P e_c\rangle_{\mathcal D}=0\). The second stirring term $\mathcal P^2X_c$ returns and reduces part of the state back to the conductive direction \(\langle e_c,\mathcal P^2e_c\rangle_{\mathcal D}=-\|\mathcal P e_c\|_{\mathcal D}^2\). And so on for the subsequent stirring terms. This iterative displacement-and-return process connects advective redistribution effects to the reduction of the tracer amplitude from \(F\) to \(G=F/(1+q^2)\).

To relate this stirring sequence to the full redistribution field \(Q\), we substitute \(X=G(e_c+Q)\) from Eq.~\eqref{eq: X field} into Eq.~\eqref{eq: X + PX = Xc} and project orthogonally to \(e_c\). Writing orthogonal projections as
\(f_\perp=f-\langle e_c,f\rangle_{\mathcal D}e_c\), this gives
\begin{equation}\label{eq: Q}
  Q=Q_1-(\mathcal PQ)_\perp.
\end{equation}
Here \(Q_1=-\mathcal Pe_c=q_1e_1\) is the first stirring pattern away from the conductive state [Eq.~\eqref{eq: X series}] with amplitude \(q_1=\|Q_1\|_{\mathcal D}\), while \((\mathcal PQ)_\perp\) describes further redistribution patterns generated in addition to $Q_1$. Since \(\langle Q,(\mathcal PQ)_\perp\rangle_{\mathcal D}=0\), taking the squared norm of Eq.~\eqref{eq: Q} gives \(q^2=q_1^2-\|(\mathcal PQ)_\perp\|_{\mathcal D}^2\). Therefore, further redistribution reduces \(q\) relative to the first-stirring scale \(q_1\). In the vanishing-flow limit, redistribution is weak and the first stirring pattern dominates to leading order \(Q\sim Q_1\). The tracer state is then well approximated by the conductive field $e_c$ plus the perturbation $e_1$. At finite flow, further stirring can generate additional redistribution patterns \(Q=\sum_{n=1}^{\infty}c_ne_n\), where the redistribution patterns \(e_n\) are orthonormal in dissipation space. 

\begin{figure}[t!]
\centering
\includegraphics[width=\columnwidth]{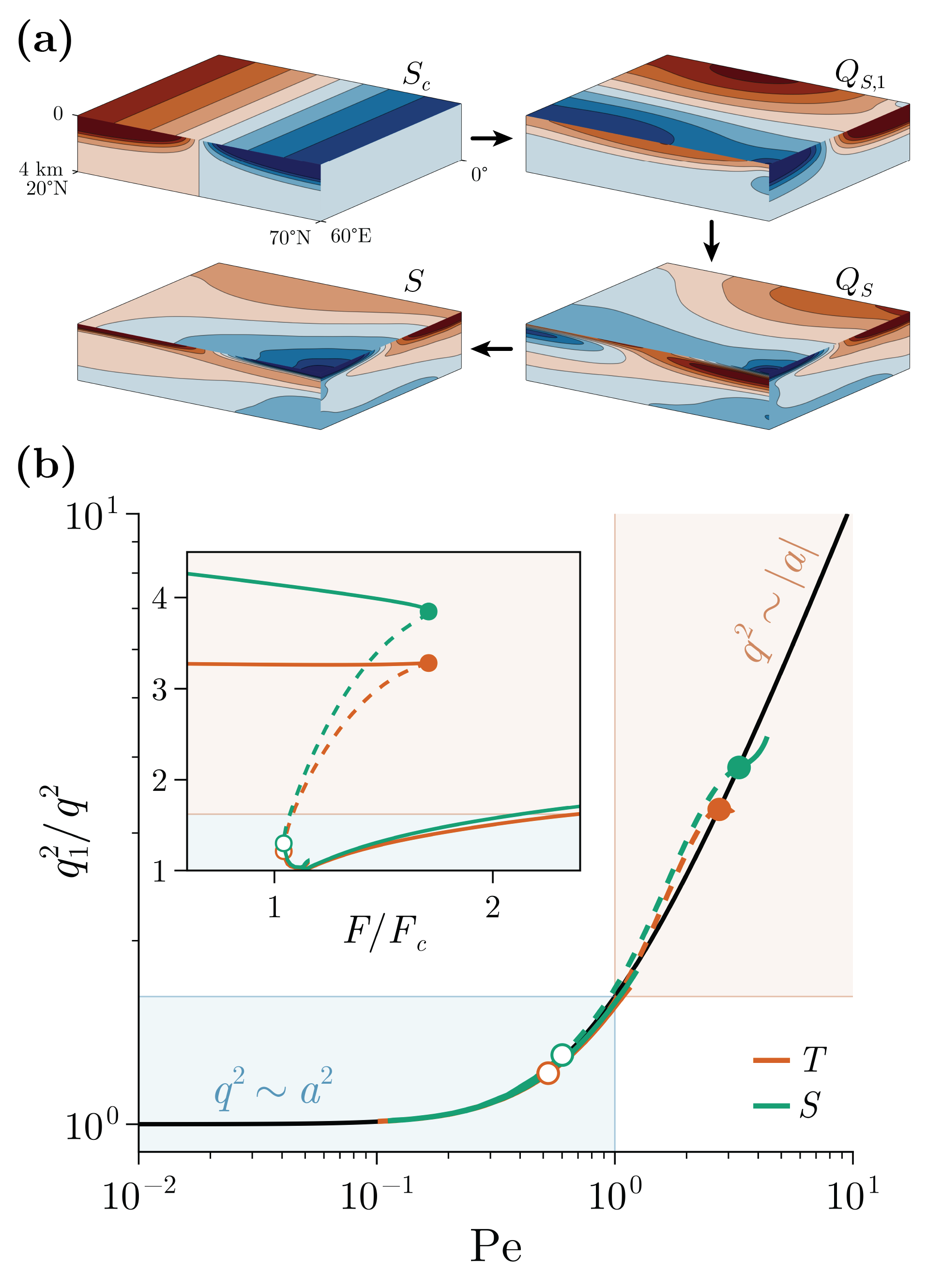}
\caption{Salinity redistribution/stirring patterns and scaling regimes in a purely buoyancy-driven thermohaline simulation with no wind-driven transport. (a) Average salinity-field components at the collapse fold: the conductive field \(S_c\), first stirring field \(Q_{S,1}\), total redistribution \(Q_S\), and realized steady state salinity field \(S\). The horizontal, latitude-depth, and longitude-depth faces show depth, zonal, and meridional averages, respectively. Each field is normalized to highlight the field structure. The arrows indicate the construction sequence [Eqs.~\eqref{eq: X series} and \eqref{eq: Q}]. (b) Transport efficiency scaling regimes for the ratio of the leading-stirring efficiency to the total efficiency $q^2_1/q^2$ versus P\'eclet number Pe [Eq.~\eqref{eq: qX(Pe)}]. No-wind simulations for temperature (orange) and salinity (green), are plotted against the P\'eclet scales \(\mathrm{Pe}_T=\hat{q}_{T,r}\) and \(\mathrm{Pe}_S=\hat{q}_{S,r}\) at the realized amplitude \(a=1\). The black curve is the prediction in Eq.~\eqref{eq: qX(Pe)}. The shaded regions and asymptotic guides identify the weak \(q^2\sim a^2\) and strong, Stommel-like \(q^2\sim\lvert a\rvert\) regimes. The inset locates the same states along the freshwater forcing bifurcation $F/F_c$. Filled and open circles mark the collapse and recovery folds, respectively. Solid branches are stable, dashed branches are unstable. }
\label{fig4}
\end{figure}

Figure~\ref{fig4}a illustrates the effects of advective redistribution for salinity at the collapse fold in a simulation without wind (to isolate buoyancy-driven circulation effects). The first stirring pattern \(Q_{S,1}\) introduces zonal contrasts absent from the conductive field \(S_c\), while further stirring generates more complex contrasts and boundary layers in the total redistribution field \(Q_S\). These complex boundary-layer structures persist in the realized salinity field \(S\), which combines the conductive $S_c$ and redistributed $Q_S$ components [Eq.~\eqref{eq: X field}]. Thus, the redistribution patterns $Q$ generated by the flow give rise to the dynamical tracer contrasts and boundary layers that are otherwise absent in the no-flow conductive state $e_c$. Importantly, the magnitude of the flow dictates the level of redistribution and in turn affects how the transport efficiency \(q=\|Q\|_{\mathcal D}\) scales with the flow strength.

To isolate how circulation strength affects transport, we consider the amplitude \(a\) of a fixed flow field, so that \(\mathcal A=a\hat{\mathcal A}\) and \(\mathcal P=a\hat{\mathcal P}\). Here, we limit ourselves to buoyancy-driven circulation without wind. The first stirring pattern has amplitude \(q_1=|a|\hat q_1\), where \(\hat q_1=\|\hat{\mathcal P}e_c\|_{\mathcal D}\) measures how the reference flow stirs the conductive pattern. For sufficiently small \(|a|\), further redistribution is weak and \(Q\simeq Q_1\), giving \(q^2\simeq q_1^2=a^2\hat q_1^2\). This quadratic dependence reflects the two stirring actions identified above: circulation generates a tracer perturbation and then transports it, with each action proportional to \(a\). A similar quadratic dependence arises in shear dispersion under rapid transverse diffusion \citep{Taylor1953}, in the thin-basin thermohaline model of \citet{CessiYoung1992}, and in a nonrotating reduction of the basin moment equations of \citet{Maas1994}.

We approximate the generation of further patterns by a self-similar redistribution hierarchy, taking the coupling between neighboring redistribution patterns to have a common strength characterized by the kinematic P\'eclet number \(\mathrm{Pe}=|a|\hat q_r\), where \(\hat q_r\) is the coupling strength (see appendix~\ref{app:stirring_hierarchy}).  This self-similar hierarchy assumption gives the approximation
\begin{equation}\label{eq: qX(Pe)}
\frac{q^2}{q_1^2}\simeq
\frac{2}{1+\sqrt{1+4\mathrm{Pe}^2}}.
\end{equation}
For \(\mathrm{Pe}\ll1\), redistribution is weak and the first-stirring pattern dominates \(q^2\simeq q_1^2\), recovering the quadratic regime above. As the flow and associated \(\mathrm{Pe}\) increase, more redistribution pathways participate and dilute the flow signal, weakening the transport efficiency $q^2$ relative to the first-stirring value \(q_1^2\). The approximate hierarchy gives \(q^2/q_1^2\sim\mathrm{Pe}^{-1}\) for \(\mathrm{Pe}\gg1\). Since both \(q_1\) and \(\mathrm{Pe}\) scale with \(|a|\), this suggests two scaling regimes
\begin{equation}\label{eq: qX(a)}
q^2(a)\sim
\begin{cases}
a^2, & \mathrm{Pe}\ll1,\\[3pt]
|a|, & \mathrm{Pe}\gg1.
\end{cases}
\end{equation}
Related transport scalings have been examined in continuously forced advection--diffusion problems for passive tracers \citep{DoeringThiffeault2006,ShawThiffeaultDoering2007}.

The relative importance of the first stirring pattern and the generation of additional redistribution patterns changes along the bifurcation branch as the forcing \(F\) varies. Figure~\ref{fig4}b compares the hierarchy approximation [Eq.~\eqref{eq: qX(Pe)}] with temperature and salinity transport along a no-wind bifurcation branch. The simulated states track the predicted relationship between \(q_1^2/q^2\) and \(\mathrm{Pe}\) [Eq.~\eqref{eq: qX(Pe)}], even as the circulation and tracer patterns change along the branch.  Near the dissipation-dominated recovery fold, the first stirring pattern provides the leading contribution to the redistribution field. For strong circulation on the ON branch, further stirring generates additional tracer patterns that modify transport and contribute to the buoyancy field driving circulation. 

\section{Box models}\label{sec:box_models}

The transport regimes above provide a direct connection to thermohaline box models. These models divide the ocean into a small number of well-mixed regions and prescribe tracer exchange laws between connected boxes using a circulation amplitude \(a\) driven by buoyancy differences between boxes. In the present framework, the corresponding advective transport is \(J_{\rm adv}=Gq^2(a)\). The linear exchange law of \citet{Stommel1961}, also used in many subsequent box models, corresponds to \(q^2\propto\lvert a\rvert\) motivated by a capillary flow analogy. \citet{Cessi1994} considered the equivalent of a quadratic exchange law $q^2\propto a^2$, motivated by the thin-basin perturbation analysis of \citet{CessiYoung1992}. The transport regimes in Eq.~\eqref{eq: qX(a)} give a physical interpretation of these choices. 
As the relative importance of the stirring patterns changes along the bifurcation branch (Fig.~\ref{fig4}b), a single prescribed exchange law may not accurately capture transport across the full branch.

To turn these tracer-transport relations into a closed system, box models make additional assumptions about how the circulation is generated. For meridionally aligned boxes, the circulation is commonly taken to scale directly with the box-scale buoyancy contrast. In a representative two-box model, temperature and salinity share the same conductive spatial pattern. With fixed normalization and geometric factors absorbed into the box-model scales, this closure takes the form
\begin{equation}\label{eq: box model}
  a\sim G_T - G_S = \frac{F_T}{1+q_T^2(a)} - \frac{F_S}{1+q_S^2(a)}.
\end{equation}
The functions \(q_T^2(a)\) and \(q_S^2(a)\) determine how a circulation of strength \(a\) transports temperature and salinity, while \(a\sim G_T-G_S\) determines how those tracer fields generate the circulation. Together they close the feedback loop and determine the steady circulation states. With the appropriate tracer-specific coefficients, Eq.~\eqref{eq: box model} recovers the steady relation of \citet{Stommel1961} when both transport efficiencies scale as \(\lvert a\rvert\), and it recovers the steady relation of \citet{Cessi1994} when they scale as \(a^2\). With $q^2\propto a^2$, Eq.~\eqref{eq: box model} also follows from a nonrotating reduction of basin-averaged angular-momentum and tracer-moment equations, assuming rapid vertical adjustment \citep{Maas2004}.

\begin{figure*}[t!]
\centering
\includegraphics[width=\textwidth]{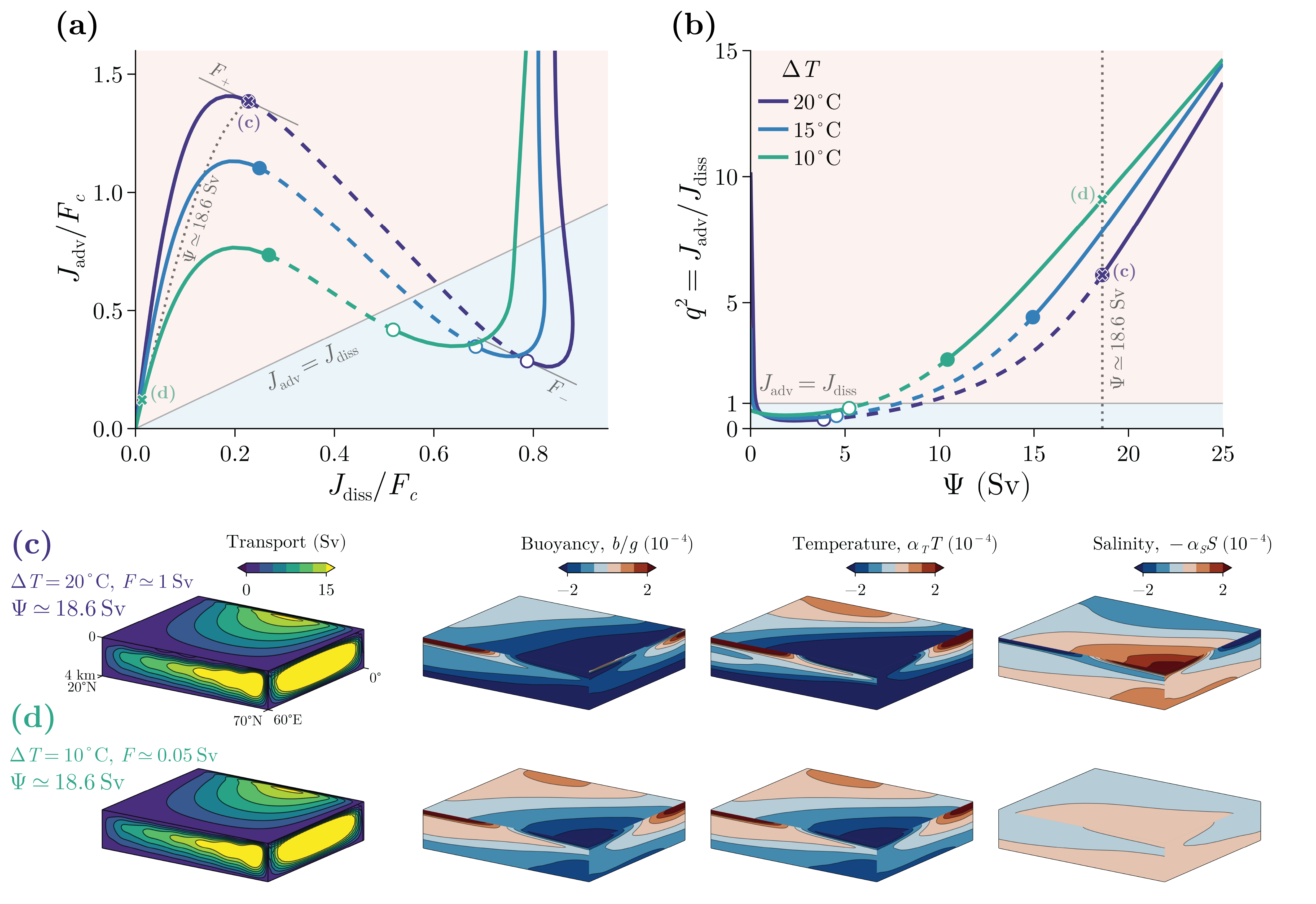}
\caption{Panel (a) shows advective \(J_{\rm adv}\) versus dissipative \(J_{\rm diss}\) salinity fluxes, both normalized by the conductive compensation scale \(F_c\) of each branch, for three surface temperature contrasts $\Delta T$ of $20\,^\circ$C (purple), $15\,^\circ$C (blue), and $10\,^\circ$C (green). At the folds (circles), \(\dd J_{\rm adv}/\dd J_{\rm diss}=-1\). The grey segments show lines of constant forcing \(F=J_{\rm adv}+J_{\rm diss}\), tangent to the \(\Delta T=20\,^\circ\mathrm C\) branch at collapse \(F_+\) and recovery \(F_-\). Background regions separate advective (red) versus dissipative (blue) dominant transport. The dotted curve interpolates between states with the same meridional overturning strength \(\Psi\simeq18.6\) Sv, with \(\Psi\) defined as in Fig.~\ref{fig1}. Panel (b) shows the same branches and information as (a) for transport efficiency $q^2=J_{\rm adv}/J_{\rm diss}$ versus overturning strength $\Psi$. Panels (c,d) show the fields at the two states marked by crosses in (a,b), both with \(\Psi\simeq18.6\) Sv. Transport streamfunctions and the depth, zonal, and meridional averages of buoyancy and its thermal and haline contributions are displayed as in Fig.~\ref{fig1}b,c. The scaled buoyancy field \(b/g\) and its thermal \(\alpha_TT\) and haline \(-\alpha_SS\) contributions share a common color scale and satisfy \(b/g=\alpha_TT-\alpha_SS\). Panel (c) shows fields at the collapse fold point $F_+\simeq 1$ Sv of the $\Delta T=20\,^\circ$C branch, whereas panel (d) shows fields far from collapse at $F\simeq 0.05$ Sv of the $\Delta T=10\,^\circ$C branch, with both states sharing the same overturning strength.}
\label{fig5}
\end{figure*}

Buoyancy-based circulation closures similar to Eq.~\eqref{eq: box model} can be derived for rotating basins by combining an approximately geostrophic interior with momentum dissipation in boundary layers \citep{WrightStockerMercer1998}. Reducing these dynamics to a relation between circulation amplitude $a$ and a single meridional buoyancy contrast (e.g., $G_T-G_S$) requires assumptions about the spatial structure of the buoyancy and flow. Changes in overturning depth, for example, can alter the relation between circulation strength and meridional buoyancy contrast \citep{deBoer2010}. Some box models explicitly account for changes in vertical structure by evolving a pycnocline depth and making the overturning depend on both depth and density contrast \citep{Johnson2007,Cimatoribus2014}. The conductive pattern \(e_c(y,z)\) (e.g., $S_c$ in Fig.~\ref{fig4}a) illustrates why the spatial structure matters. In a geostrophic ocean basin, meridional buoyancy gradients support zonal thermal-wind shear. This purely zonal geostrophic flow $\bm u_{\mathrm{geo}}=(u_{\mathrm{geo}},0,0)$ does not directly redistribute the zonally uniform conductive tracer $e_c(y,z)$, \(\bm u_{\mathrm{geo}}\boldsymbol{\cdot}\boldsymbol\nabla e_c =u_{\mathrm{geo}}\partial_xe_c=0\). Inferring meridional overturning from this zonal shear requires additional assumptions about pressure gradients and friction near the basin boundaries. These pressure gradients depend on the full buoyancy field, including the redistributed temperature and salinity represented by \(Q\). The redistribution fields in Fig.~\ref{fig4}a illustrate the zonal tracer contrasts and boundary layers that contribute to this buoyancy field. They therefore affect both tracer transport and the buoyancy gradients that drive the circulation. The closure \(a\sim G_T-G_S\) assumes that tracer contrasts along the meridional direction remain sufficient to determine the circulation despite changes in the redistributed buoyancy field. That proportionality is expected to hold when both tracer balances are dissipation dominated ($q^2<1$), as near the recovery fold, because the weak flow ensures that tracer fields remain near their conductive patterns $e_c$ with amplitudes $G$ (Fig.~\ref{fig4}b). It may not remain valid where the salinity balance is advection dominated ($q^2>1$), as along the ON branch approaching collapse (Fig.~\ref{fig3}b), because the redistributed tracer fields and boundary layers in $Q$ then set much of the buoyancy structure that drives the circulation [Eq.~\eqref{eq: b field}]. 

Despite their approximations, box models retain the key salt-advection feedback dynamics under mixed boundary conditions that reproduce the saddle-node bifurcation geometry found in more complex models \citep{Weijer2019}. Calibrated box models have also reproduced GCM hysteresis \citep{Rahmstorf1996} and transient responses \citep{Schleussner2011}, and estimated collapse thresholds \citep{Wood2019}. However, fitted parameters may compensate for omitted processes, so fitted agreement alone does not establish whether the prescribed transport and circulation laws remain valid as the forcing and tracer structure change. While the closure problem remains complex for low-order, box-model representations of the system, the basin-scale flux balance [Eq.~\eqref{eq: F_X = G_X(1+q_X^2)}] connects prescribed inter-box exchange laws to tracer transport in the continuous system [Eqs.~\eqref{eq: T,S PDEs}]. This gives us a way to test prescribed box-model exchange laws, such as \(q^2\propto|a|\), against the tracer transport and circulation strength measurements along a bifurcation branch in spatially resolved simulations.

\section{Implications and Discussion}\label{sec:discussion}

The flux balance $F=J_{\rm diss}+J_{\rm adv}=G(1+q^2)$ derived here provides a basin-scale constraint on the thermohaline ocean state. Common diagnostics such as sea-surface-temperature anomalies \citep{Caesar2018} and meridional overturning strength \citep{Johns2011} are used to assess thermohaline circulation and its response to changing forcing. The MOC strength $\Psi$, for example, measures the volumetric transport of water, but how much heat and salt are transported also depends on how these tracers are distributed throughout the ocean basin. Knowing $\Psi$ alone therefore does not specify the tracer transport that balances the imposed forcing, or establish how close the circulation is to collapse. 

Figure~\ref{fig5} highlights these implications explicitly. It shows simulated freshwater-forcing bifurcation branches for three different atmospheric temperature contrasts $\Delta T$. We identified two distinct states at different freshwater forcings on separate $\Delta T$ branches, with the same overturning strength $\Psi\simeq 18.6$ Sv, yet one state occurs exactly at the collapse fold $F_+$ and the other occurs far from collapse. These two states share similar circulation patterns but the tracer and buoyancy fields that drive the circulation are distinct (Fig.~\ref{fig5}c,d). And when comparing transport efficiencies $q^2$ to $\Psi$ across different branches we do not find a unique relation organizing their connection, with notably different $q^2$ values at the two identified $\Psi\simeq 18.6$ Sv states (Fig.~\ref{fig5}b). The flux balance $F=J_{\rm diss}+J_{\rm adv}$, by contrast, is able to distinguish these states by their different advective $J_{\rm adv}$ and dissipative $J_{\rm diss}$ flux contributions. In addition, tipping points occur where $F$ reaches a maximum (collapse) or minimum (recovery) along a branch, so that $\dd F=\dd J_{\rm diss}+\dd J_{\rm adv}=0$ (e.g., Fig.~\ref{fig2}e). Small increases in dissipative flux are then compensated by equal decreases in advective flux, $\dd J_{\rm adv}=-\dd J_{\rm diss}$, so the branch slope in the $J_{\rm adv}$--$J_{\rm diss}$ plane, which reaches $-1$ at the folds, also distinguishes the two states' trends toward tipping (Fig.\ref{fig5}a). This clarifies the limitations of using $\Psi$ as a proxy for thermohaline state and distance to tipping.

The flux balance framework separates relations that follow exactly from tracer conservation from those that require approximation. At any steady state of a closed domain, the flux balance $F=J_{\rm diss}+J_{\rm adv}=G(1+q^2)$ [Eq.~\eqref{eq: F_X = G_X(1+q_X^2)}] holds exactly, with $G$ and $q$ defined directly from the realized and conductive tracer fields, and the tracer field taking the form $X=G(e_c+Q)$ with $\|Q\|_{\mathcal D}=q$ [Eq.~\eqref{eq: X field}]. These relations require no closure for the circulation. By contrast, the estimate $F_-\sim F_c$ is a scale argument based on conductive compensation, and the transport regimes $q^2\propto a^2$ and $q^2\propto|a|$ follow from the self-similar redistribution hierarchy [Eq.~\eqref{eq: qX(Pe)}], an approximation that the no-wind simulations broadly follow (Fig.~\ref{fig4}b). Box models then correspond to prescribed exchange laws $q^2(a)$ combined with a circulation closure such as $a\sim G_T-G_S$ [Eq.~\eqref{eq: box model}], each of which may hold in one transport regime but not across a full bifurcation branch.

Several assumptions limit the present results. The flux balance applies to closed domains; for regional domains such as the Atlantic basin, heat and salt transports through the open boundaries must be accounted for, as in transport diagnostics such as the overturning-based freshwater flux through the southern boundary (often written as $F_{\rm ov}$) which has been shown to contain dynamical information about bistability \citep{Rahmstorf1996,deVriesWeber2005}. Another limitation is that along each simulated branch, the surface-forcing pattern and the dissipation operator are fixed. In coupled models, however, air--sea feedbacks alter the surface fluxes \citep{RahmstorfWillebrand1995,Vanderborght2025}, and parameterized mixing depends on the ocean state \citep{Griffies2000}, so applying the framework there requires deciding how these state-dependent effects enter the forcing and dissipation operators. Hosing experiments with a localized freshwater anomaly may also need separate consideration (section~\ref{sec:flux_balance} and appendix~\ref{app:background_anomaly}). The framework also describes equilibria, so transient experiments must vary slowly enough for the system to remain in quasi-steady state. Finally, our simulations use an idealized single-hemisphere basin with a quasi-steady diagnostic velocity and equal momentum and tracer diffusivities. The framework does not depend on these choices, but the transport regimes and the estimate $F_-\sim F_c$ should be tested in models with richer dynamics. Slowly forced hysteresis experiments in GCMs \citep{vanWestenDijkstra2023,Hawkins2011} provide a natural next test for the flux balance, in which $G$ and $q$ can be diagnosed along the equilibrium branches both from the total forcing and with the passive tracer of appendix~\ref{app:background_anomaly}, followed eventually by ocean state estimates that close the heat and salt budgets \citep{Forget2015}.

Thus, the framework outlined here in Part~I of this work provides basin-scale thermohaline constraints as well as a physical interpretation of the bifurcation dynamics of this system. Namely, while circulation and tracer patterns vary with forcing, the net effects of these changes can be expressed through advective and dissipative transport contributions that combine to balance the forcing. This formulation connects box-model and related thermohaline circulation work to the continuous tracer conservation equations, and provides a starting point for constraining these dynamics in more complex models such as GCMs. In Part~II, we extend the framework by developing and testing approximate closures for the flux balance to forecast tipping points from measurements of
pre-tipping states.  

\acknowledgments
The authors would like to acknowledge funding support from the National Science Foundation (NSF) under award OCE 2343203.

\availstatement
Model code and output will be made available upon publication.

\appendix[A]
\appendixtitle{Numerical model}
\appendixlabel{A}{app:numerical_model}

\subsection{Governing equations}

The simulations solve Eqs.~\eqref{eq: T,S PDEs} in the rectangular basin described in section~\ref{sec: numerics}, with $0\le x\le W$, $0\le y\le L$, and $-H\le z\le 0$. Temperature and salinity diffuse with the momentum viscosity coefficients, $\bm K_T=\bm K_S={\rm diag}(\nu_h,\nu_h,\nu_v)$. Surface fluxes enter as forcing and restoring terms distributed over a thin surface layer,
\begin{equation}
 \mathcal F_T-\mathcal R_TT=\gamma_T\,h_0(T_*-T),\qquad
 \mathcal F_S=h_0\,Q_S,
 \label{eq:app_forcing}
\end{equation}
where $h_0(z)=e^{z/\delta_0}/[\delta_0(1-e^{-H/\delta_0})]$ integrates to one over the depth. We retain a finite forcing depth $\delta_0=0.05H=200$~m \citep[e.g.,][]{Soons2025}, which we interpret as a simple representation of unresolved near-surface redistribution. Smoothing the imposed fluxes over this depth also assists modal convergence of the surface-forced fields. 
In dimensional units, the atmospheric temperature is $T_*=\tfrac12\Delta T\cos(\pi y/L)$ and the salt flux is $Q_S=(\pi S_0F/WL)\cos(\pi y/L)$, with reference salinity $S_0=35$~psu, so that $F$ is the total freshwater input to the northern half of the basin.

The velocity is diagnosed from the momentum equations without the time-derivative and advection terms \citep[e.g.,][]{Huck1999},
\begin{align}
 f\hat{\bm z}\times\bm u&=-\bm\nabla p+b\,\hat{\bm z}+\nu_h\nabla_h^2\bm u+\nu_v\partial_{zz}\bm u -r\bm u+h_0\bm\tau,\nonumber\\
 \bm\nabla\cdot\bm u&=0,
 \label{eq:app_momentum}
\end{align}
where $p$ is the pressure divided by the reference density $\rho_0$, $b=g(\alpha_T T-\alpha_SS)$ is the dimensional buoyancy, $r=10^{-6}$~s$^{-1}$ is the Rayleigh drag coefficient \citep[e.g.,][]{Huck1999}, and $f=2\Omega\sin(\varphi_0+y/R)$ for Earth's rotation rate $\Omega$ and radius $R$, with $\varphi_0=20^\circ$ at the southern boundary. 
The zonal wind stress $\rho_0\bm\tau=-\tau_0\cos(\pi y/L)\hat{\bm x}$ is westward in the south and eastward in the north.
The walls are impermeable and free slip. Without a time derivative, the velocity adjusts instantly to the buoyancy field and wind stress, giving the flow $\bm u(b)$ introduced with Eqs.~\eqref{eq: T,S PDEs}.


\subsection{Diagnostic velocity}

We write the velocity in terms of toroidal and poloidal potentials $\psi$ and $\phi$ \citep[e.g.,][]{Chu2003},
\begin{equation}
 \bm u=\hat{\bm z}\times\bm\nabla\psi+\bm\nabla\times(\hat{\bm z}\times\bm\nabla\phi),
 \label{eq:app_potentials}
\end{equation}
which satisfies $\bm\nabla\cdot\bm u=0$ identically. The toroidal part carries the horizontal circulation, with vertical vorticity $\zeta=\nabla_h^2\psi$, and the poloidal part carries the overturning, with vertical velocity $w=\nabla_h^2\phi$. This reduces the number of fields by two (needing only $\psi$ and $\phi$ to resolve the three velocity fields and pressure field). Taking the vertical component of the curl and of the double curl of Eq.~\eqref{eq:app_momentum} eliminates the pressure and gives one equation for each potential,
\begin{align}
 \beta v-f\partial_zw&=(\nu_h\nabla_h^2+\nu_v\partial_{zz}-r)\,\zeta-h_0\partial_y\tau,\nonumber\\
 \partial_z(f\zeta-\beta u)&=\nabla_h^2b+(\nu_h\nabla_h^2+\nu_v\partial_{zz}-r)\nabla^2w,
 \label{eq:app_vorticity}
\end{align}
where $\beta=\partial_yf$, and $u$ and $v$ are the zonal and meridional velocities. This system remains nonhydrostatic through the frictional term $(\nu_h\nabla_h^2+\nu_v\partial_{zz}-r)\nabla_h^2w$, where the remaining hydrostatic component is $(\nu_h\nabla_h^2+\nu_v\partial_{zz}-r)\partial_{zz}w$. Without friction and wind, the first equation is the planetary vorticity balance $\beta v=f\partial_zw$, and the second reduces to the divergence of the thermal-wind balance that follows from geostrophic and hydrostatic balance. Viscosity and drag close the circulation in boundary currents along the walls.

In bounded domains, this velocity transformation couples the boundary conditions of the two potentials, $\psi$ and $\phi$ \citep{BoronskiTuckerman2007}. Taking the double curl of Eq.~\eqref{eq:app_momentum} leaves a residual boundary term on the lateral walls. Retaining this term adds thin wall flows that produce oscillations and poor convergence in the global Fourier modes used to solve the fields (described below). We therefore omit this extra boundary term and solve Eqs.~\eqref{eq:app_vorticity} as an interior balance, consistent with other interior geostrophic models \citep[e.g.,][]{Huck1999}. The omitted wall term leaves the kinetic energy budget technically unclosed. However, the velocity remains divergence free with no flow through the walls, so tracer conservation is unaffected.

\subsection{Numerical solution}

We expand $\psi$, $\phi$, $T$, and $S$ in Fourier modes that each satisfy the boundary conditions. With $\tilde x=x/W$, $\tilde y=y/L$, and $\tilde z=(z+H)/H$, the Fourier basis functions have the forms
\begin{align}
 \psi&:\ \sin(m\pi\tilde x)\sin(n\pi\tilde y)\cos(\ell\pi\tilde z),\nonumber\\
 \phi&:\ \cos(m\pi\tilde x)\cos(n\pi\tilde y)\sin(\ell\pi\tilde z),\nonumber\\
 T,S&:\ \cos(m\pi\tilde x)\cos(n\pi\tilde y)\cos(\ell\pi\tilde z),
 \label{eq:app_basis}
\end{align}
with $m,n\ge1$ and $\ell\ge0$ for $\psi$, $m,n\ge0$ and $\ell\ge1$ with $(m,n)\ne(0,0)$ for $\phi$, and $m,n,\ell\ge0$ for the tracers. Each function is normalized to unit volume-averaged squared magnitude. Each velocity mode is impermeable and free slip at every boundary, and each tracer mode has no diffusive flux through the boundaries. We project each equation onto its own basis \citep[Galerkin projection; e.g.,][]{deville2002high} to produce a system of ODEs evolving Fourier amplitude vectors for each tracer, with diagnostic linear matrix equations for the toroidal and poloidal velocity potential amplitudes.

The projected tracer equations form the dynamical system written schematically as
\begin{align}
 \dot{\bm T}&=-\big[\bm D+\bm R_T+\bm A(\bm b)\big]\bm T+\bm F_T,\nonumber\\
 \dot{\bm S}&=-\big[\bm D+\bm A(\bm b)\big]\bm S+\bm F_S,
 \label{eq:app_odes}
\end{align}
for the vectors of tracer amplitudes $\bm T$ and $\bm S$, with $\bm b=\bm T-\bm S$ (in dimensionless form). Here $\bm D$ is the diagonal diffusion matrix, $\bm R_T$ and $\bm F_T$ represent thermal restoring and forcing, $\bm F_S$ is the projected salinity forcing (proportional to the freshwater forcing amplitude $F$), and $\bm A(\bm b)$ is the advection matrix of the diagnosed flow. Because each velocity mode is divergence free and impermeable, $\bm A$ is antisymmetric ($\bm A^{\top}=-\bm A$) for any truncation, ensuring tracer conservation, so the flux balance [Eq.~\eqref{eq: F = J^diss + J^adv}] holds at the simulated equilibria.

We concentrate tracer resolution in the vertical direction to represent thin surface boundary layers that extend over broad horizontal scales. For each horizontal mode $(m,n)$, we retain vertical modes up to a maximum that decreases with the larger of $m$ and $n$, so we omit modes with both fine horizontal and fine vertical structure. Specifically, the tracers retain $0\le m,n\le12$ horizontal modes and all vertical modes $0\le\ell\le\ell_{\max}$, where $\ell_{\max}=\lfloor10+70(2/7)^{\max(m,n)/3}\rfloor$ decreases exponentially from 80 at the largest horizontal scales to 10 at the smallest. This gives 2746 coefficients per tracer, compared with 13689 for the full rectangular set. The velocity retains the rectangular sets $1\le m\le60$, $1\le n\le26$, and $0\le\ell\le60$ for $\psi$, and $0\le m\le60$, $0\le n\le25$, and $1\le\ell\le60$ with $(m,n)\ne(0,0)$ for $\phi$, about $10^5$ coefficients each. The larger number of zonal velocity modes resolves the western boundary current in the momentum inversion. 


We follow stable and unstable equilibrium branches in $F$ by pseudo-arclength continuation using \texttt{BifurcationKit.jl} \citep{Veltz2020}, starting from equilibria obtained by implicit time integration after slowly spinning up the system from zero tracer and momentum surface forcing. Linear stability is assessed from the Jacobian of the coupled tracer equations [Eqs.~\eqref{eq:app_odes}], including the response of the diagnosed flow to tracer perturbations.

\appendix[B]
\appendixtitle{Localized forcing anomalies}
\appendixlabel{B}{app:background_anomaly}

When the freshwater forcing varies in amplitude with a fixed pattern and the dissipation is fixed, as in our simulations, $e_c$ is fixed along each branch and the framework applies as is. In GCM hosing experiments, however, a freshwater anomaly $\mathcal F'$ concentrated in the North Atlantic is added to a background forcing $\bar{\mathcal F}$ with a different pattern \citep[e.g.,][]{Jackson2023}; here we treat $\bar{\mathcal F}$ as fixed. The flux balance [Eq.~\eqref{eq: F_X = G_X(1+q_X^2)}] still holds with $F$ and $e_c$ computed from the total conductive field $\bar S_c+S_c'$. However, $e_c$ then changes with the anomaly amplitude, and in a global domain the balance can be dominated by transport in other basins, which may change little as the Atlantic overturning collapses. The transport regimes and compensation scale (section~\ref{sec:bistability}) and the closures of Part~II may then need to be reconsidered. Here we outline one way to retain a fixed conductive pattern.

Consider a passive tracer $S'$ that is forced only by the anomaly and transported by the realized flow and mixing, but does not affect buoyancy:
\begin{equation}
 \mathcal D_SS'+\mathcal A(b)S'=\mathcal F'.
 \label{eq:app_passive}
\end{equation}
Because tracer transport is linear for a given flow, $S-S'$ is the background salinity carried by the same flow. Thus $S'$ is not the change from a control state, which also includes background salinity redistributed by the changed circulation. With the anomaly conductive field $S_c'=F'e_c'$, where $\mathcal D_SS_c'=\mathcal F'$ and $\|e_c'\|_{\mathcal D_S}=1$, the derivation of section~\ref{sec:flux_balance} applied to Eq.~\eqref{eq:app_passive} gives
\begin{equation}
 F'=G'(1+q'^2),
 \label{eq:app_passive_balance}
\end{equation}
with
\begin{equation}
 G'=\langle e_c',S'\rangle_{\mathcal D_S},\quad
 q'^2=\frac{\|S_c'-S'\|^2_{\mathcal D_S}}{\|S'\|^2_{\mathcal D_S}}.
 \label{eq:app_passive_balance G q}
\end{equation}
The conductive pattern $e_c'$ is fixed along a hosing branch, and $q'$ measures the transport of the anomaly pattern only. When the background and anomaly share a pattern, $q'=q$ and $G'=(F'/F)G$. Because Eq.~\eqref{eq:app_passive_balance} has the form of the main-text balance, the fold and cusp constraints (Part~II, section~3) hold in $(F',G',q')$ wherever $G'$ labels the equilibria, and the closures can be posed in these variables. At the control state, where $F'=0$, $q'$ follows from the passive response to a unit anomaly. In a GCM, $S'$ can be carried as an additional passive tracer forced by the hosing flux, as done for added heat in FAFMIP \citep{Gregory2016}.

Along a hosing branch, the conductive buoyancy is $b_c=T_c-\bar S_c-F'e_c'$. Minimizing $\|b_c\|_{\mathcal D_S}$ over the anomaly amplitude, as for Eq.~\eqref{eq: Fc}, gives
\begin{equation}
 F'_c=\langle e_c',T_c-\bar S_c\rangle_{\mathcal D_S},
 \label{eq:app_Fc_anomaly}
\end{equation}
so the background conductive salinity joins $T_c$ in the buoyancy that the anomaly compensates. Since $\mathcal D_Se_c'=\mathcal F'/F'$, Eq.~\eqref{eq:app_Fc_anomaly} is equivalent to $\langle\mathcal F',b_c\rangle=0$ at $F'=F'_c$: weighted by the anomaly flux, the conductive buoyancy is the same where freshwater is added and where it is removed. Thus $F'_c$ depends on where the compensating removal is placed. Unlike $T_c$ and $S_c$ in our simulations, which have similar spatial structures, a localized anomaly can compensate only part of the background conductive buoyancy, and the remainder can still drive circulation.

The passive form may retain much of the structure of Part~II and of the compensation scale, but this must be tested in hosing experiments with localized anomalies, alongside other constructions such as the total balance above.

\appendix[C]
\appendixtitle{Self-similar redistribution hierarchy}
\appendixlabel{C}{app:stirring_hierarchy}

Here we derive Eq.~\eqref{eq: qX(Pe)} by following how further stirring changes the first response $Q_1=q_1e_1$. We hold the realized flow fixed when applying $\mathcal P=\mathcal D^{-1}\mathcal A(b)$ and restrict $\mathcal D$ to its positive-definite space, removing the conserved salinity mean. Taking the inner product of $Q=Q_1-(\mathcal PQ)_\perp$ [Eq.~\eqref{eq: Q}] with $Q$ gives
\begin{equation}
 q^2=\langle Q,Q_1\rangle_{\mathcal D}=q_1c_1,
 \label{eq:app_first_pattern}
\end{equation}
since $\langle Q,(\mathcal PQ)_\perp\rangle_{\mathcal D}=0$. Here $c_1=\langle Q,e_1\rangle_{\mathcal D}$ is the amplitude of the first pattern remaining in the realized redistribution field. Determining how further stirring changes $c_1$ therefore determines $q^2$.

Stirring $e_1$ returns part of the response to $e_c$ and generates a new pattern $e_2$, with coupling strength $q_2=\|(\mathcal Pe_1)_\perp\|_{\mathcal D}$. Repeating this construction, removing earlier components and normalizing each new pattern, gives
\begin{align}
 \mathcal Pe_1&=q_1e_c-q_2e_2,\nonumber\\
 \mathcal Pe_n&=q_ne_{n-1}-q_{n+1}e_{n+1},\qquad n\ge2.
 \label{eq:app_lanczos}
\end{align}
The patterns are orthonormal in the dissipation-weighted inner product, and $q_{n+1}=\|\mathcal Pe_n-q_ne_{n-1}\|_{\mathcal D}$. Antisymmetry, $\langle f,\mathcal Pg\rangle_{\mathcal D}=-\langle\mathcal Pf,g\rangle_{\mathcal D}$, leaves only neighboring couplings in this constructed basis. The $q_n$ measure coupling strengths, whereas the $c_n$ in $Q=\sum_{n\ge1}c_ne_n$ measure the realized pattern amplitudes.

Projecting Eq.~\eqref{eq: Q} onto these patterns gives
\begin{align}
 c_1+q_2c_2&=q_1,\nonumber\\
 c_n-q_nc_{n-1}+q_{n+1}c_{n+1}&=0,\qquad n\ge2.
 \label{eq:app_chain}
\end{align}
Eliminating successive amplitudes and using $q^2=q_1c_1$ yields
\begin{equation}
 q^2=\cfrac{q_1^2}{1+\cfrac{q_2^2}{1+\cfrac{q_3^2}{1+\cdots}}}.
 \label{eq:app_cfrac}
\end{equation}
In the finite-dimensional model, this fraction terminates when a new coupling vanishes and is exact for the fixed flow. Keeping only the first pattern ($q_2=0$) gives $q^2=q_1^2$.

We now approximate all subsequent couplings by the common strength $q_n=|a|\hat q_r=\mathrm{Pe}$ for $n\ge2$ and extend the hierarchy indefinitely. Here $\mathcal P=a\hat{\mathcal P}$ scales a fixed flow shape, where $a$ represents the buoyancy-driven flow amplitude with no wind-driven effects. The repeating fraction satisfies $q^2/q_1^2=[1+\mathrm{Pe}^2(q^2/q_1^2)]^{-1}$, whose positive solution is
\begin{equation}
 \frac{q^2}{q_1^2}=\frac{2}{1+\sqrt{1+4\mathrm{Pe}^2}},
 \label{eq:app_selfsimilar}
\end{equation}
recovering Eq.~\eqref{eq: qX(Pe)}. Weak stirring gives $q^2\simeq q_1^2\propto a^2$; strong stirring distributes the response among many patterns, giving $q^2\simeq q_1^2/\mathrm{Pe}\propto|a|$. The latter scaling depends on the infinite common-coupling approximation. In Fig.~\ref{fig4}b, we set $\hat q_r=q_2$ at the realized amplitude $a=1$, measuring $\mathrm{Pe}$ from the flow and conductive pattern without using the realized $q$.

This continued fraction is structurally analogous to representations of effective diffusivity in passive-tracer stirring \citep{AvellanedaMajda1991,MajdaKramer1999}. For constant isotropic diffusion without restoring, $\mathrm{Nu}=1+q^2$ is the square of the gradient-based mixing efficiency of \citet{DoeringThiffeault2006} and \citet{ShawThiffeaultDoering2007}. Their bounds for smooth sources in periodic domains with statistically homogeneous, isotropic stirring permit linear and quadratic growth of $\mathrm{Nu}$ with flow amplitude, consistent with Eq.~\eqref{eq: qX(Pe)}.

\bibliographystyle{ametsocV6}
\bibliography{references}

\end{document}